\documentclass[%
reprint,
superscriptaddress,
 amsmath,amssymb,
 aps,
]{revtex4-1}

\usepackage{graphicx}
\usepackage{dcolumn}
\usepackage{bm}
\usepackage{braket}
\usepackage{xcolor}

\begin{document}

\title{Theory of Cation Solvation and Ionic Association in Non-Aqueous Solvent Mixtures}

\author{Zachary A. H. Goodwin}
\email{zachary.goodwin13@imperial.ac.uk}
\affiliation{Department of Materials, Imperial College of London, South Kensington Campus, London SW7 2AZ, UK}
\affiliation{John A. Paulson School of Engineering and Applied Sciences, Harvard University, Cambridge, MA 02138, USA}

\author{Michael McEldrew}
\affiliation{Department of Chemical Engineering, Massachusetts Institute of Technology, Cambridge, MA, USA}

\author{Boris Kozinsky}
\affiliation{John A. Paulson School of Engineering and Applied Sciences, Harvard University, Cambridge, MA 02138, USA}
\affiliation{Robert Bosch LLC Research and Technology Center, Cambridge, MA 02139, USA}

\author{Martin Z. Bazant}
\affiliation{Department of Chemical Engineering, Massachusetts Institute of Technology, Cambridge, MA, USA}
\affiliation{Department of Mathematics, Massachusetts Institute of Technology, Cambridge, MA, USA}

\begin{abstract}
Conventional lithium-ion batteries, and many next-generation technologies, rely on organic electrolytes with multiple solvents to achieve the desired physicochemical and interfacial properties. The complex interplay between these properties can often be elucidated via the coordination environment of the cation. We develop a theory for the coordination shell of cations in non-aqueous solvent mixtures that can be applied with high fidelity, up to super-concentrated electrolytes. Our theory can naturally explain simulation and experimental values of cation solvation in ``classical'' non-aqueous electrolytes. Moreover, we utilise our theory to understand general design principles of emerging classes of non-aqueous electrolyte mixtures, such as high entropy electrolytes. It is hoped that this theory provides a systematic framework to understand simulations and experiments which engineer the solvation structure and ionic associations of concentrated non-aqueous electrolytes.
\end{abstract}

\date{\today}

\maketitle

\section{Introduction}

Lithium-ion batteries (LIBs) have transformed our everyday lives through myriad portable electronic devices~\cite{xu2004nonaqueous,xu2014electrolytes,Zheng2017Uni,li2020new,Tian2021,Kang2022Nav,Piao2022,Cheng2022Sol}. Although current LIBs are tremendously successful, further development is necessary to meet safety and performance requirements for widespread use of electric vehicles and grid-level energy storage of renewable energy~\cite{li2020new,Tian2021,Kang2022Nav,Piao2022,Cheng2022Sol}. Thus, much of electrochemical energy storage research is focused on improving the materials beyond commercial LIBs, including high voltage and high capacity cathode materials~\cite{li2020high}, silicon anodes~\cite{chen2020electrolyte}, lithium metal anodes~\cite{wang2022liquid,yu2020molecular,yu2022rational}, and even non-lithium based battery chemistries~\cite{qin2019localized,zheng2018extremely,chen2021highly}. A requirement of any next-generation battery chemistry is an electrolyte that can operate safely, efficiently and reversibly over the course of the battery's lifetime~\cite{xu2004nonaqueous,xu2014electrolytes,Zheng2017Uni,li2020new,Tian2021,Kang2022Nav,Piao2022,Cheng2022Sol}.

One of the primary design concepts in electrolyte engineering is balancing desired, often seemingly conflicting, physicochemical properties, such as low viscosity and high conductivity, with interfacial properties, such as low charge transfer resistance and electrode passivation~\cite{li2020new}. Commercial LIBs accomplish this design concept, to some extent, with blends of linear carbonates, such as dimethyl carbonate (DMC) for low viscosity, and cyclic carbonates, such as ethylene carbonate (EC) to passivate carbonaceous anodes~\cite{xu2004nonaqueous,xu2014electrolytes,Zheng2017Uni,li2020new,Tian2021,Kang2022Nav,Piao2022,Cheng2022Sol,Xie2023}. Although carbonate blends provide a serviceable solution, electrolyte design for LIBs is far from optimized~\cite{li2020new,Tian2021,Kang2022Nav,Piao2022,Cheng2022Sol}.

An important advance in electrolyte engineering over the past decade has been the link of ion solvation chemistry to electrolyte properties~\cite{xu2007solvation,von2012correlating,borodin2020uncharted,andersson2020ion,Cheng2022Sol,Wang2021corsol,Kim2021Pot}. On one hand, at low salt concentrations, solvent dominated coordination shells ensure a nearly complete dissociation of salt, enabling efficient ion transport. However, even at low concentrations, when multiple types of solvent are used, a cation's solvation structure can be tuned via the bulk solvent ratios to produce very different solid-electrolyte interphases (SEIs) to passivate the working electrodes~\cite{von2012correlating}. This is in large part due to the increased reactivity, from reduced LUMO levels, of solvent molecules that are in direct contact with lithium. At higher salt concentrations, the introduction of anions to the coordination shell tends to decrease ionic conductivity and increase concentration overpotentials~\cite{xu2004nonaqueous,xu2014electrolytes}. However, in some cases, highly passivating inorganic anion-derived (as opposed to solvent-derived) SEIs can form, resulting in an overall increase in cell performance~\cite{Suo2013}. Furthermore, across all salt concentrations and solvent ratios, the solvation/desolvation dynamics governing charge transfer resistance will be subject to large changes depending on the species comprising the coordination shells of cations in the electrolyte~\cite{Cheng2022Sol,Piao2022}.

In fact, next-generation electrolytes are engineered to tune the solvation shell of cations, and therefore physicochemical and interfacial properties of the electrolytes~\cite{li2020new,Tian2021,Kang2022Nav,Piao2022,Cheng2022Sol,Chen2021Steric,yu2022rational}. For example, localized high-concentration electrolytes (LHCE)~\cite{chen2018high,zheng2018extremely,chen2021highly,ren2018localized,qin2019localized,cao2021localized,Liu2022BS} use typical salts, such as fluoro-sulfonamides, in high ratios with ether solvents which historically showed promising compatibility with lithium metal anodes, such as 1,2-dimethoxyethane, but which are notoriously limited in terms of oxidative stability~\cite{koch1978stability,koch1982specular,foos1988new,qian2015high}. The singular aspect of LHCE's is their use of diluents, generally in the form of fluorinated ethers, which are non-solvating, but highly oxidatively stable~\cite{chen2018high,zheng2018extremely,chen2021highly,ren2018localized,qin2019localized,cao2021localized,Liu2022BS}. Thus, the non-fluorinated solvent and anion are mostly retained in the lithium coordination shell, where it has enhanced oxidative stability, while the fluorinated ether diluent strictly appears outside of the lithium coordination shell, serving to lubricate the electrolyte (decrease viscosity) while remaining stable against both electrodes~\cite{chen2018high,zheng2018extremely,chen2021highly,ren2018localized,qin2019localized,cao2021localized,Liu2022BS}. However, a major issue in engineering such next-generation electrolytes is knowing exactly how the solvation shell of the cations will be effected by the \textit{mixtures} which are created to balance the physicochemical and interfacial properties, as they are often not just an average of their neat properties.

To probe the solvation structure of electrolytes, spectroscopic techniques~\cite{xu2004nonaqueous,xu2014electrolytes,Cheng2022Sol} such as Fourier-transform infrared~\cite{barthel2000ftir}, Raman~\cite{Cresce2017,seo2012electrolyte1,seo2012electrolyte2}, nuclear magnetic resonance~\cite{yang2010investigation} and mass spectrometry~\cite{von2012correlating,Zhang2018ISI}, have been employed to give some insight. However, atomistic simulations, including molecular dynamics (MD)~\cite{Postupna2011,von2012correlating,Borodin2014SEI,Li2015,Skarmoutsos2015,Borodin2017Mod,Han2017MD,Ravikumar2018,Shim2018,Han2019MD,Piao2020Count,Hou2021,Wu2022} and density functional theory (DFT)~\cite{Cresce2017,Borodin2017Mod,Beltran2020LHCE,Hou2021,Wu2022}, are typically relied on to give detailed insight into ion solvation. An advantage of DFT is predicting HOMO/LUMO levels of different species, and therefore, their expected reactivity; in addition to understanding the spectra from different techniques~\cite{Cresce2017}. However, DFT is computationally expensive, and often limited to short length and time scales (when performing \textit{ab initio} MD)~\cite{Beltran2020LHCE}. To reach larger scales, MD simulations are indispensable in understanding solvation structures and dynamics of ions but the accuracy and fidelity of the force field often comes under question~\cite{Postupna2011,von2012correlating,Borodin2014SEI,Li2015,Skarmoutsos2015,Borodin2017Mod,Han2017MD,Ravikumar2018,Shim2018,Han2019MD,Piao2020Count,Hou2021,Wu2022}. Recent advances in machine learning force fields could offer the accuracy of DFT with the time/length scales of classical force fields~\cite{Ioan2022,Yao2022CRev}. However, a systematic way to report solvation structures has not emerged, with practically every study reporting results in a different way, making it difficult to compare different works systematically~\cite{Kang2022Nav}. Moreover, atomistic simulations can make predictions of specific systems, but they are not able to provide a general understanding that a simple, analytical theory often does.

In this paper, we develop a theory for the solvation structure and ionic associations of concentrated non-aqueous electrolytes. This theory is based on McEldrew \textit{et al.}'s~\cite{mceldrew2020theory,mceldrew2020corr,mceldrew2021ion,mceldrew2020salt} recent generalisation of  Flory~\cite{flory1941molecular,flory1941molecular2,flory1942constitution,flory1942thermodynamics,flory1953principles,flory1956statistical}, Stockmayer~\cite{stockmayer1943theory,stockmayer1944theory,stockmayer1952molecular} and Tanaka's~\cite{tanaka1989,tanaka1990thermodynamic,tanaka1994,ishida1997,tanaka1995,tanaka1998,tanaka2002,tanaka1999,tanaka2011polymer} work on reversible aggregation and gelation in polymer systems to concentrated electrolytes. Our theory only has a handful of physically transparent parameters, all of which can be obtained directly from atomistic simulations. To demonstrate this parameterization, we perform MD simulations of a ``classical'' non-aqueous electrolyte, 1M~LiPF$_6$ in a mixture of ethylene carbonate and propylene carbonate and compare the results to our theory. Overall, our theory is able to naturally explain the solvation structure of this electrolyte for all the studied mixtures. Having benchmarked our theory against a well studied system, we utilise ``toy models'' of our theory to understand the design principles of emerging next-generation LIB electrolytes. The theory is able to provide a depth of understanding, which is indispensable in aiding design principles of these electrolytes. Moreover, the theory provides a systematic framework for simulation techniques to report their results, which we hope will be adopted to bring more transparency to the field. 

\section{Theory}

The central idea that the presented work is based on, is the analogy between concentrated electrolytes and polymers. In concentrated electrolytes, past the Kirkwood transition~\cite{Kirkwood1936}, electrostatic correlations between ions causes the aggregation of oppositely charged ions into structures with alternating spatial order of oppositely charged ions~\cite{mceldrew2020theory,levy2019spin}. Conceptually, this is similar to the formation of an alternating co-polymer of two monomers. The realisation of this analogy actually has a long history, with one of the first connections being the similarity between the liquid-vapour phase diagram of ions in their high temperature gas state and the phase diagram of polymers in solution, as first discussed from a general stand
point by Pitzer~\cite{Pitzer1990,Pitzer1991book}. Simulations of the phase behavior of ions in their high temperature dissociated state, based on the so-called restricted primitive model~\cite{Fisher1994}, have directly revealed the formation of chains of ions~\cite{Fisher2002,Yan2002}, although this effect has not been quantified in relation to understanding the phase boundary shape. The resemblance of these phase diagrams is thought to occur because of the similarity between polymers and fluids of dipolar particles~\cite{Workum2006,Dudowicz2004} to self-assemble into dynamic polymer chains (``equilibrium polymers"), and also between dipolar fluids and ionic fluids, owing to the association of ions into dipolar and multipolar clusters~\cite{Shelley1995,Workum2006}. There are also many examples of the observation of chains, rings and branched ionic aggregates, amongst other morphologies, forming in simulations of electrolytes~\cite{Bastea2002,Yan2002,Fisher2002,Doye1999,Yu2022Agg,choi2014ion,choi2015ion,choi2015ion3,choi2017ion}. We note that the polymeric chains of ions have been found in simulations
of ion clusters that are thought to be important in biomineralization~\cite{Demichelis2011}. 

Here, as in our previous works~\cite{mceldrew2020salt,mceldrew2020theory,mceldrew2021ion,mceldrew2020corr}, we develop and adapt the theories of Flory-Stockmayer-Tanaka for thermoreversible polymerisation and apply it to LIB electrolytes, where solvent molecules play an essential role in the molecular aggregation behaviour, further establishing the analogy between concentrated electrolytes and polymers. In fact, this is not the first time that inspiration from polymer-based models has been used to further deepen the analogy between electrolytes and polymers. A notable example comes from Bastea~\cite{Bastea2002}, who developed a simple theory of the formation of chains and rings comprising of alternating cations/anions that was observed in MD simulations of a highly size asymmetric electrolyte. The inspiration for Bastea's theory for these structures came from models for worm-like surfactant micelles~\cite{Cates1990}, with Bastea assuming the monomer unit was a neutral ion pair. 

The connection between Flory-Stockmayer-Tanaka theories has already been made with ``patchy particle'' or ``spot model'' systems, where a particle has a fixed number of patches/spots that can favourably interact with the patches/spots of other particles, forming thermoreverisble associations~\cite{kumar2001gelation,saika2004effect,zaccarelli2005model,zaccarelli2007colloidal,Russo2009,corezzi2009connecting,Smallenburg2013,Bianchi2008,Bianchi2007,Sciortino2007,Audus2016,Audus2018}. Similarly, Stockmayer fluids, particles interacting with Lennard-Jones potentials with additional directional potentials, such as dipolar (or higher order) interactions, have also been shown to form thermoreversible associations~\cite{Workum2006,Dudowicz2004}. These simulation approaches are not limited to the mean-field approximation with Cayley trees, but the mean-field theories of Flory-Stockmayer-Tanaka with Cayley trees have successfully been used to understand the aggregation and gelation of patchy colloidal particles~\cite{groschel2013,Russo2009,corezzi2009connecting,Bianchi2008,Bianchi2007,Sciortino2007}. Recently, systems of oppositely charged (patchy) nanoparticles were synthesis and mixed together, forming alternating aggregates of positively and negatively charged nanoparticles~\cite{Kalsin2006,Kalsin2007,Umar2013}. 

Here, we develop a theory for a ``classical'' non-aqueous electrolyte - 1M LiPF$_6$ dissolved in a mixture of ethylene carbonate (EC) and propylene carbonate (PC) - because of its importance in LIBs~\cite{xu2004nonaqueous,xu2014electrolytes}. While this is used as an example here, the theory is not limited to specifically this system, and can be adapted for a wide variety of electrolytes, up to extremely high salt concentrations. 

It is assumed that the cations bind with the anions and solvent molecules, but no other species bind with each other, i.e. no binding between anion and the solvents (although this can be taken into account~\cite{mceldrew2020theory}) or between solvent molecules (which has been considered elsewhere~\cite{choi2018}). This non-aqueous electrolyte is considered to form a polydisperse mixture of ionic clusters of rank $lmsq$, containing $l$ cations, $m$ anions, $s$ EC solvent molecules and $q$ PC solvent molecules, of which there are $N_{lmsq}$~\cite{mceldrew2020theory}. 

The cations can form a maximum of $f_+$ associations and the anions a maximum of $f_-$ associations, referred to as their functionality. The solvent molecules are assumed to have a functionality of 1. The cations and anions are able to form extended clusters (because $f_\pm > 2$ for LiPF$_6$)~\cite{Yu2022Agg}. However, as the concentration of the salt is relatively low and the interactions between the solvents and Li cations strong, a percolating ionic network (gel) should not form at ambient conditions~\cite{mceldrew2020theory}. Therefore, for simplicity, we neglect the formation of the gel for 1M LiPF$_6$ in EC/PC, and refer readers to Refs.~\citenum{mceldrew2020theory,mceldrew2020corr,mceldrew2021ion,mceldrew2020salt} to see how it is included. 

\begin{figure}
    \centering
    \includegraphics[width = 0.49\textwidth]{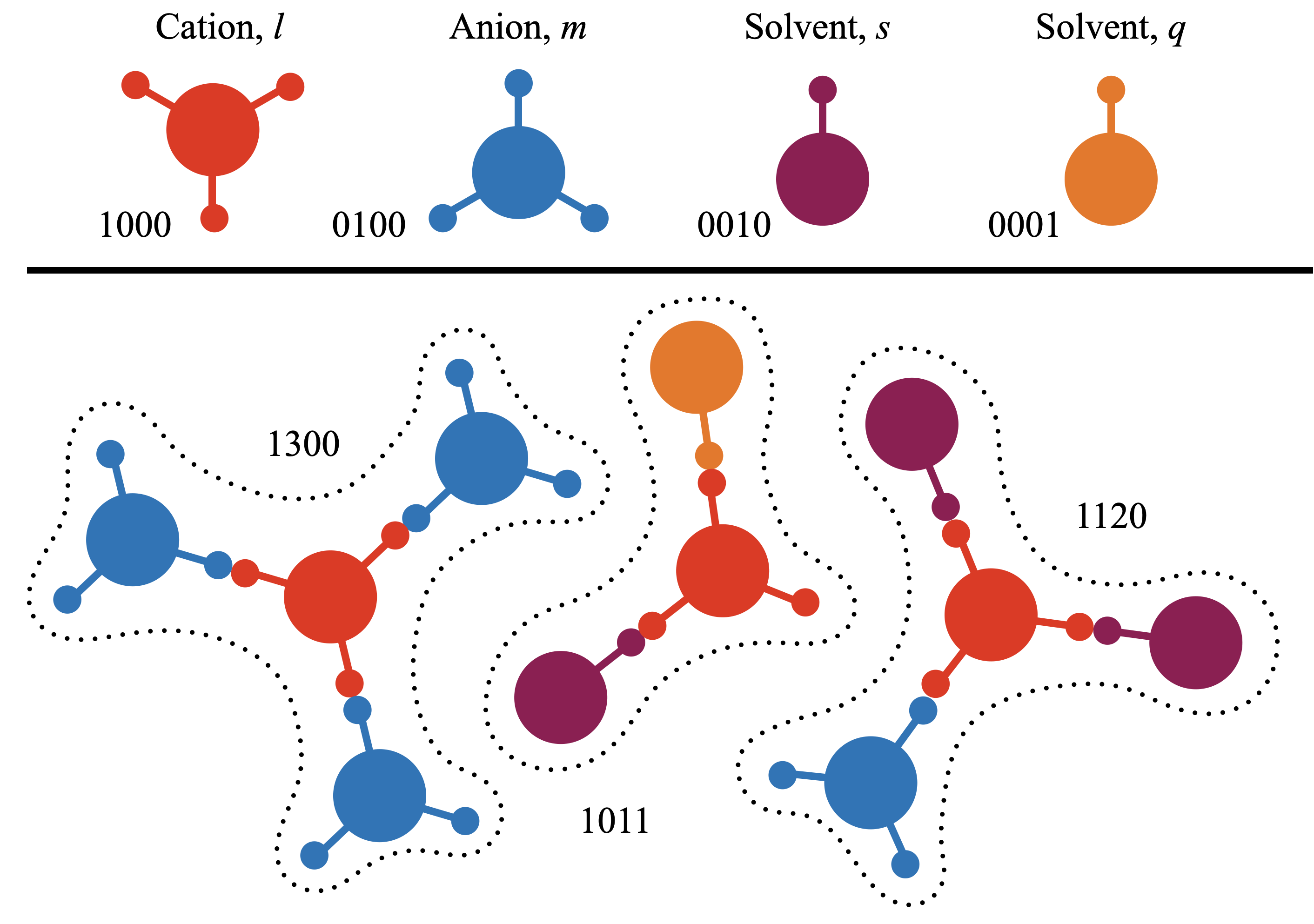}
    \caption{Schematic showing the notation of different species. The upper panel shows cations, anions, and the two solvents, with overall quantities denoted by $+,-,x,y$, respectively, and individual species in clusters by $l,m,s,q$, respectively. Free cations, anions and $x$ and $y$ solvents are denoted by the rank labels $1000$, $0100$, $0010$ and $0001$, respectively. In the bottom panel, example clusters are shown, where dotted lines are used to enclose each cluster.}
    \label{fig:clusters}
\end{figure}

The electrolyte is assumed to form Cayley tree clusters~\cite{mceldrew2020theory, mceldrew2020salt, Goodwin2022GEL}. The cations and anions form the ``backbone of the branched aggregates'', and the ``dangling association sites'' from this backbone can either be left empty or can be decorated with the solvent molecules, or the strong solvent-cation interactions can break apart the ionic backbone. These Cayley tree clusters have no intra-cluster loops, with the number of each type of bond being uniquely determined by the rank of the cluster, $lmsq$. 

In Fig.~\ref{fig:clusters} a schematic of some example clusters for the studied system are displayed, where the cations and anions have a functionality of $3$. The cluster of rank $1300$ is comprised of $1$ cation and $3$ anions, which has an overall $-2$ negative charge, cannot bind to any solvent molecules from the lack of open cation association sites. For $1011$, the cation is bound to both solvent molecules, and there is a dangling association site which could bind to an anion or another solvent molecule to create a cluster of a larger rank. Finally, the cluster of rank $1120$ is a cation-anion ion pair, where the cation is also bound to two of the $s$ solvent molecules. These are just some examples of cluster to demonstrate the possible allowed clusters. 

The assumption of Cayley tree clusters is an essential requirement of the presented theory, enabling simple analytical results, but it is important to note its potential limitations. This assumption only appears to work if there is at least one species that is disorder-forming. If there are only order-forming species, then the ionic aggregates that form will be crystalline in nature and have a significant number of loops~\cite{choi2018}. For example, molten/concentrated NaCl only contains spherical, highly symmetric ions, both of which are order forming, and the resulting clusters that form a crystalline-like~\cite{choi2018,Doye1999}. If, however, the anion is a fluoro-sulfonamide, the aggregates which form are no longer crystalline-like, but do in fact resemble Cayley trees, as we have previously shown~\cite{mceldrew2021ion,mceldrew2020corr,mceldrew2020salt}. Therefore, the presented theory is not applicable to all electrolytes.

A natural question to ask is if this assumption works for LIB electrolytes. In ``classical'' non-aqueous electrolytes the salt is often LiPF$_6$, which might be thought to be order-forming species, as both cation and anion have high symmetry. In the context of salt-in-ionic liquids, we have shown that the Cayley tree assumption can actually be used to represent the small clusters formed in that system~\cite{mceldrew2020salt}. Moreover in Ref.~\citenum{Xie2023}, 4~M LiPF$_6$ in EC was studied from MD simulations and experiments, and it was shown that branched, Cayley-tree-like aggregates form. In the context of next-generation electrolytes, disordered anions are often used due to their high solubility, which suggests that our theory should be applicable to these systems~\cite{Kartha2021}. Moreover, the carbonate solvents are not highly symmetric species and strongly interact with the cation, which should break up any large ionic aggregates in favour of small solvation structures. Therefore, we believe the assumption of Cayley tree clusters should be justified for most LIB electrolytes.

In the presented theory, there are $N_+$ Li$^+$ cations and $N_-$ [PF$_6$]$^-$ anions, and from electroneutrality in the bulk $N_+ = N_-$. This salt is dissolved in a mixture of EC and PC, of which there are $N_x$ and $N_y$, respectively. As described above and shown in Fig.~\ref{fig:clusters}, these species can form a polydisperse mixture of clusters of different ranks, of which there are $N_{lmsq}$. The theory assumes the electrolyte to be an incompressible lattice fluid (this is not an essential requirement, but is taken for convenience, and does not change any results here), with a single lattice site having the volume of a Li cation, $v_+$~\cite{mceldrew2020theory}. Anions occupy $\xi_- = v_-/v_+$ lattice sites, and the solvents occupy $\xi_{x/y} = v_{x/y}/v_+$ lattice sites. Note that the volumes of each species only explicitly appear in the volume fractions, and the different shapes of species are not explicitly accounted for. The total number of lattice sites is given by
\begin{align}
    \Omega=\sum_{lmsq}(l+\xi_-m+\xi_xs+\xi_yq)N_{lmsq}.
    \label{eq:O}
\end{align}

\noindent Dividing through by the total number of lattice sites gives
\begin{equation}
    1 = \sum_{lmsq}(l+\xi_-m+\xi_xs+\xi_yq)c_{lmsq} = \sum_{lmsq}\phi_{lmsq}.
\end{equation}

\noindent Here $c_{lmsq}=N_{lmsq}/\Omega$ is the dimensionless concentration of a $lmsq$ cluster (the number of $lmsq$ clusters per lattice site), and $\phi_{lmsq}$ is the volume fraction of clusters of rank $lmsq$. The volume fraction of each species is determined through
\begin{equation}
    \phi_i = \sum_{lmsq}\xi_i j c_{lmsq},
\end{equation}

\noindent where $j = l,m,s,q$ for $i = +,-,x,y$, respectively, and $\xi_+ = 1$. The incompressibility condition can also be stated as
\begin{align}
    1 = \phi_++\phi_-+\phi_x+\phi_y.
    \label{eq:incomp}
\end{align}



Based on the works of Flory~\cite{flory1941molecular,flory1941molecular2,flory1942constitution,flory1942thermodynamics,flory1953principles,flory1956statistical}, Stockmayer~\cite{stockmayer1943theory,stockmayer1944theory,stockmayer1952molecular} and Tanaka~\cite{tanaka1989,tanaka1990thermodynamic,tanaka1994,ishida1997,tanaka1995,tanaka1998,tanaka2002,tanaka1999,tanaka2011polymer}, and our previous application of this theory to concentrated electrolytes~\cite{mceldrew2020theory,mceldrew2020corr,mceldrew2021ion,mceldrew2020salt,Goodwin2022GEL,Goodwin2022IP}, the free energy is taken to be
\begin{equation}
\mathcal{F} = \sum_{lmsq} \left[N_{lmsq}k_BT\ln \left( \phi_{lmsq} \right)+N_{lmsq}\Delta_{lmsq}\right],
\label{eq:FSI}
\end{equation}

\noindent where the first term is the ideal entropy of each cluster of rank $lmsq$, and the second term is the free energy of forming those clusters, with $\Delta_{lmsq}$ denoting the free energy of formation for a cluster of rank $lmsq$\cite{mceldrew2020theory}, which has three contributions
\begin{align}
\Delta_{lmsq}=\Delta_{lmsq}^{bind}+\Delta_{lmsq}^{comb}+\Delta_{lmsq}^{conf},
\label{eq:Delta}
\end{align}

\noindent where $\Delta_{lmsq}^{bind}$ is the binding energy, $\Delta_{lmsq}^{comb}$ is the combinatorial entropy, and $\Delta_{lmsq}^{conf}$ is the configurational entropy. 

The binding energy of a cluster, $\Delta^{bind}_{lmsq}$, is uniquely defined by the rank $lmsq$, owing to the assumption of Cayley tree clusters. For $l > 0$, the binding energy of a cluster is
\begin{align}
    \Delta^{bind}_{lmsq}=(l+m-1)\Delta u_{+-} + s\Delta u_{+x} + q\Delta u_{+y},
    \label{eq:dbond}
\end{align}

\noindent where $\Delta u_{ii'} = \Delta u_{i'i}$ is the energy of an association between $i$ and $i'$. When $l = 0$, the binding energy of a cluster is $0$.

Physically, the cation-anion associations are driven by their strong electrostatic interactions~\cite{mceldrew2020theory}. These associations are a representation of short-range electrostatic correlations beyond mean-field~\cite{mceldrew2020theory,mceldrew2020corr,mceldrew2021ion,mceldrew2020salt,Goodwin2017,Chen2017,feng2019free}, with Levy \textit{et al.}~\cite{levy2019spin} showing that only retaining short-ranged correlations in concentrated systems is sufficient to reproduce the spatial arrangements of concentrated electrolytes. The fact that we have neglected long-ranged electrostatic interactions \textit{between different clusters}, relies on the assumption that the majority of the electrostatic energy is incorporated in the formation of the ionic clusters (via this binding contribution), as opposed to between different clusters. Moreover, the electrolyte is at a concentration where the activity coefficient of the salt is typically larger than $1$~\cite{xu2014electrolytes}, which motivates the exclusion of further electrostatic energy terms, with our theory predicting such values of activity coefficients for concentrated electrolytes~\cite{mceldrew2020theory,mceldrew2021ion}. The cation-solvent associations are driven by the strong interactions between Li and the carbonyl oxygen. It is assumed that cation-solvent interactions are the dominant contribution to the free energy of solvation, instead of, say, the Born solvation energy.

%
%

The combinatorial entropy of each cluster of rank $lmsq$ is given by
\begin{equation}
    \Delta_{lmsq}^{comb} = - k_BT\ln\left\{f_+^lf_-^mW_{lmsq}\right\},
\end{equation}

\noindent where 
\begin{equation}
    W_{lmsq} = \dfrac{(f_+l -l)!(f_-m - m)!}{l!m!s!q!(f_+l - l - m - s - q +1)!}.
\end{equation}

\noindent This entropic term was first derived by Stockmayer for applications in polymers~\cite{stockmayer1952molecular}. It only depends on the number of each species in a cluster, since it is a measure of the number of ways in which a cluster of rank $lmsq$ can be arranged.

The configurational entropy is associated with building the aggregate from its constituent parts on a lattice, which depends on the volumes of each species. This contribution is not universal, and can depend on how ``fexible'' the associations are~\cite{mceldrew2020corr}. Importantly, this contribution, in the end, produces a term that depends on the number of associations in a cluster, which means it contributes to the \textit{entropy change of an association}, as seen in the association constant later. The binding energy and configurational entropy could, therefore, be combined into a single binding free energy, but for historical reasons we have not described the theory in this way. We refer the reader to Ref.~\citenum{mceldrew2020corr} for further details of the configurational entropy, where the temperature dependence was studied.


Establishing chemical equilibria, as shown in Ref.~\citenum{mceldrew2020theory}, the dimensionless concentration of a cluster of rank $lmsq$, with $l > 0$, is given by 
\begin{align}
     c_{lmsq}=\frac{W_{lmsq}}{\Lambda_{-}}\left(\psi_{l}\Lambda_{-} \right)^{l} \left(\psi_{m}\Lambda_{-} \right)^{m}(\psi_{s}\Lambda_{x})^s(\psi_{q}\Lambda_{y})^q,
    \label{eq:dist}
\end{align}

\noindent where $\psi_{l} = f_+\phi_{1000}/\xi_+$ and $\psi_{m} = f_+\phi_{0100}/\xi_+$ are number of association sites per lattice site for free cations and free anions, respectively, and $\psi_{s} = \phi_{0010}/\xi_x$ and $\psi_{q} = \phi_{0001}/\xi_y$ are number of association sites per lattice site for the free solvent $x$ and $y$, respectively. The association constant between species $i$ and cations is given by
\begin{align}
    \Lambda_{i}=e^{-\beta\Delta f_{+i}}=e^{-\beta\left[\Delta u_{+i} - T\Delta s_{+i}\right]},
\end{align}

\noindent where $\beta$ is inverse thermal energy, and $\Delta s_{+i}$ is the entropy change of an association, as determined by the configurational entropy~\cite{mceldrew2020theory,mceldrew2020corr}. Note $i \neq +$ in these equations, as no cation-cation associations are considered. When $l = 0$ and $m = 1$ (or $m = 0$ and $l = 0$), and $s = q = 0$, Eq.~\eqref{eq:dist} yields the correct limit for free anions (cations). For $l = 0$, the free volume fractions of $x$ and $y$ are separately defined, as $\Delta^{bind}_{0010} = \Delta^{bind}_{0001} = 0$, which is not a limit of Eq.~\eqref{eq:dbond}.

Here, the associations constants are the main parameters of the theory that need to be determined. As shown in the next section, and in our previous works~\cite{mceldrew2020corr,mceldrew2020salt,mceldrew2021ion}, these parameters of the theory are readily determined from the ensemble average coordination numbers. Therefore, the theory does not contain any free parameters which can be arbitrarily fitted. In principle, as has been utilised in patchy particle systems, the Wertheim theory can be employed to have a parameter-free model~\cite{Wertheim1984I,Wertheim1984II,Wertheim1986III,Wertheim1986IV,Teixeira2019,Bianchi2007,Sciortino2007,Liu2009,Audus2016,Audus2018}. This theory requires that a reference hard sphere pair correlation function and the Mayer function for the interacting species is known. As we are applying the theory to LIB electrolytes, which contain complex molecular species, we believe using the association constants as parameters to be found by performed MD simulations, and using the theory to understand the results of these simulations in more depth, is more appropriate than making \textit{a priori} predictions for these electrolytes.

The volume fractions of each \textit{free} species is (usually) not a known input, however. In fact, these quantities are often what we aim to determine from the theory. To overcome this seemingly circular problem, the free volume fraction of each species can be related to the total volume fraction of that species, the number of bonds that species can make to the other species, and the probability of the associations~\cite{mceldrew2020theory}. The probability of species $i$ being associated to species $j$ is given by $p_{ij}$, the number of associations was previously defined as the functionality, and the total volume fractions of each species is known. For free cations $\phi_{1000}= \phi_+(1-\sum_ip_{+i})^{f_+}$, for free anions $\phi_{0100}=\phi_-(1-p_{-+})^{f_-}$, and for free solvent we have $\phi_{0010}=\phi_x(1-p_{x+})$ and $\phi_{0001}=\phi_y(1-p_{y+})$.

The association probabilities are related through the conservation of associations and mass action laws. The conservation of associations states
\begin{align}
\psi_+ p_{+i} = \psi_i p_{i+} = \Gamma_{i},
\label{eq:sys1}
\end{align}

\noindent where $\Gamma_i$ is the number of $+i$ associations per lattice site, and $\psi_i=f_i \phi_i/\xi_i=f_ic_i$ is the number of association sites of species $i$ per lattice site. Again, note $i \neq +$ in these equations, which means for the 4-component system being discussed, there are 3 sets of these equations. Similarly, the set of 3 mass action laws for the associations are generally given by
\begin{align}
    \Lambda_{i}\Gamma_{i} = \frac{p_{i+}p_{+i}}{(1-\sum_{i'}p_{+i'})(1-p_{i+})}.
    \label{eq:sys2}
\end{align}

\noindent By solving the set of equations described by Eqs.~\eqref{eq:sys1} and \eqref{eq:sys2}, one can determine the association probabilities, and therefore the volume fraction of free species and the cluster distribution. For the 4-component system, numerical solutions are required. For the assumption that there is no gel, $(f_+ - 1)(f_- - 1) > p_{+-}p_{-+}$ must hold, as this inequality determines the percolation point on a Bethe lattice~\cite{mceldrew2020theory}.

In the context of polymers, Tanaka~\cite{tanaka1989} introduce the mass action laws to describe the thermorevrsible nature of the bonds. In electrolytes, the use of mass action laws to describe the equilibrium association has a long history, starting from Arrhenius and Bjerrum pairs~\cite{marcus2006ion}. Since those early works, it has been employed countless times to model ion pairing effects~\cite{marcus2006ion}. Therefore, the theory presented here is an extension of these works, where we are able to account for much larger ionic aggregates in a systematic way, in addition to including solvation effects, and not requiring aggregates to be neutral.

Having outlined the basic equations, before studying the example of non-aqueous electrolytes in detail, we believe it is prudent that some intuition is outlined from limiting cases. It is suggested that the reader also goes through our general theory~\cite{mceldrew2020theory}, and special cases for ionic liquids~\cite{mceldrew2020corr,Goodwin2022GEL,Goodwin2022IP}, water-in-salt electrolytes~\cite{mceldrew2021ion} and salt-in-ionic liquids~\cite{mceldrew2020salt}, as these contain a lot of details of the ionic associations not shown here. Let us take $\Delta f_{+i} = 0$ for all $i$, which would correspond to $\Lambda_i = 1$. In such a case, despite the free energy of an association being zero, the association probabilities are finite, which means there is a distribution of clusters of various ranks. This occurs because of the ideal entropy of mixing and the combinatorial entropy each cluster drives the formation of more types of clusters with more different components in them, as this increases the entropy~\cite{mceldrew2020theory,Goodwin2022GEL}. We know, however, from Flory’s lattice expression for the entropy of disorientation~\cite{flory1942thermodynamics,flory1953principles}, that $\Delta s_{+i} > 0$. At low temperature, the negative binding energies dominate, and $\Lambda_i > 1$, but at high temperatures the entropy term dominates and $\Lambda_i \approx 0$. In this latter limit, the clusters dissociate, and we are left with only free species, with the ideal entropy of each species reducing down to the entropy of mixing of different species on a lattice~\cite{mceldrew2020theory,Goodwin2022GEL}.

\section{Cation Solvation in Classical Non-Aqueous Electrolytes}

To test our theories ability to understand cation solvation, we perform atomistic MD simulations using LAMMPS~\cite{plimpton1995,LAMMPS2022} of LiPF$_6$ in mixtures of EC and PC. For each simulation, we have 83 Li cations and 83 PF$_6$ anions. In addition, there are (83, 167, 250, 333, 417, 500, 583, 667, 750) EC and (750, 667, 583, 500, 417, 333, 250, 167, 83) PC solvent molecules. The initial configurations for all simulations were generated using PACKMOL~\cite{martinez2009packmol}. An initial energy minimisation was performed before equilibrating the bulk, periodic system. The production run used to analyse the associations in this system were run at 300~K in an NVT ensemble (the volume of the simulation box was determined from NPT equilibration at 1~bar) for 10~ns with a 1~fs time step. 

We employed the CL$\&$P force field for Li cations and PF$_6$ anions~\cite{lopes2012}, with the van der Waals radius of Li cations set to $1.44~\textrm{\AA}$~\cite{Ravikumar2018}. For EC and PC, the OPLS/AA force fields were utilised~\cite{Jorgensen1996,Jorgensen2005,Dodda2017,Dodda2017NUC}, which are commonly used for simulations of these electrolytes. Long range electrostatic interactions were computed using the particle-particle particle-mesh solver (with a cut-off length of 12~$\textrm{\AA}$).

It has been shown numerous times, and as found here, that the pair correlation function (or cumulative coordination number) of Li-O (carbonyl), has a large peak (sharp increase) at $\sim2~\textrm{\AA}$, with a pronounced minimum (long plateau) at $\sim3~\textrm{\AA}$ until a moderate second peak at $4-5~\textrm{\AA}$~\cite{Ravikumar2018,Han2017MD,Han2019MD}. The Li-O (ether), in contrast, has its first large peak at $\sim3~\textrm{\AA}$. Therefore, it is natural to define a EC/PC solvent molecule to be in the coordination shell if the Li-O (carbonyl) separation is less than 2.5~$\textrm{\AA}$. Whereas, the Li-F pair correlation function (cumulative coordination number) often has much smaller peak(s) (much steadier increase) at $3-4~\textrm{\AA}$, despite the van der Waals radius of F (in the employed MD simulation) only being only slightly larger than O~\cite{Ravikumar2018,Shim2018,Han2017MD,Han2019MD}. Therefore, we shall focus solely on the solvents in the cation coordination shell, not delving into the information about cation-anion association in this section.

\begin{figure}[h]
\centering
\includegraphics[width=0.49\textwidth]{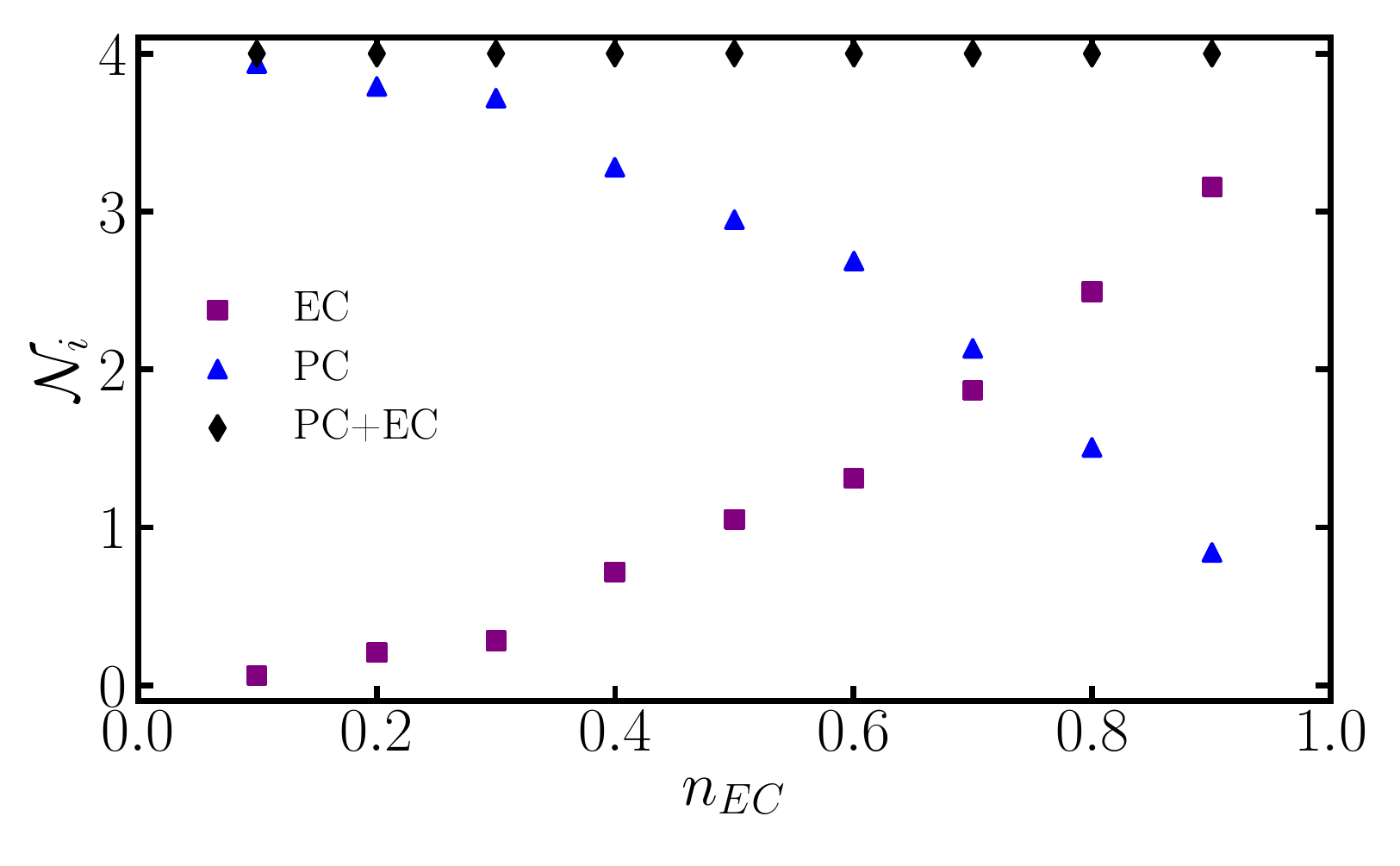}
\caption{Number of EC/PC molecules associated to Li, $\mathcal{N}_i$, as a function of the mole fraction of EC solvent to the total mole fraction of solvent, $n_{EC}$.}
\label{fig:simulations_N}
\end{figure}

Using this cut-off distance to define an association between Li and EC/PC, we computed the ensemble average coordination number of Li cations, $\mathcal{N}_i$ where $i$ is one of these solvent molecules, for the various mixtures of EC/PC. The results of this simulation analysis are shown in Fig.~\ref{fig:simulations_N}, plotted as a function of the mole fraction of EC to the total mole fraction of solvent, $n_{EC}$. We find that the $\mathcal{N}_{EC}$ increases monotonically with $n_{EC}$ and $\mathcal{N}_{PC}$ decreases monotonically with $n_{EC}$, as might be expected from an increasing proportion of EC. Their dependence on $n_{EC}$ is non-linear, but the sum of the solvents in the coordination shell of Li remains to be $\sim4$ for all $n_{EC}$. 

The composition of the first coordination shell of Li is a quantity which is often calculated from simulations. It provides insight into the interactions in the system, and is often correlated with certain behaviours of the electrolyte, such as the conductivity or SEI forming abilities. Moreover, within our theory, the ensemble average coordination numbers is a key quantity in determining the association probabilities, $\mathcal{N}_i = f_+p_{+i}$, i.e. the functionality (maximum coordination number) of Li multiplied by the probability that Li is associated to $i$ gives the coordination number of that species. In Fig.~\ref{fig:simulations_N}, it is clear that $f_+ = 4$. Therefore, the association probabilities of cations binding to other species, $p_{+i}$, can be readily determined, and from Eq.~\eqref{eq:sys1} the association probabilities of species binding to Li, $p_{i+}$, can be found.

\begin{figure}[h]
\centering
\includegraphics[width=0.49\textwidth]{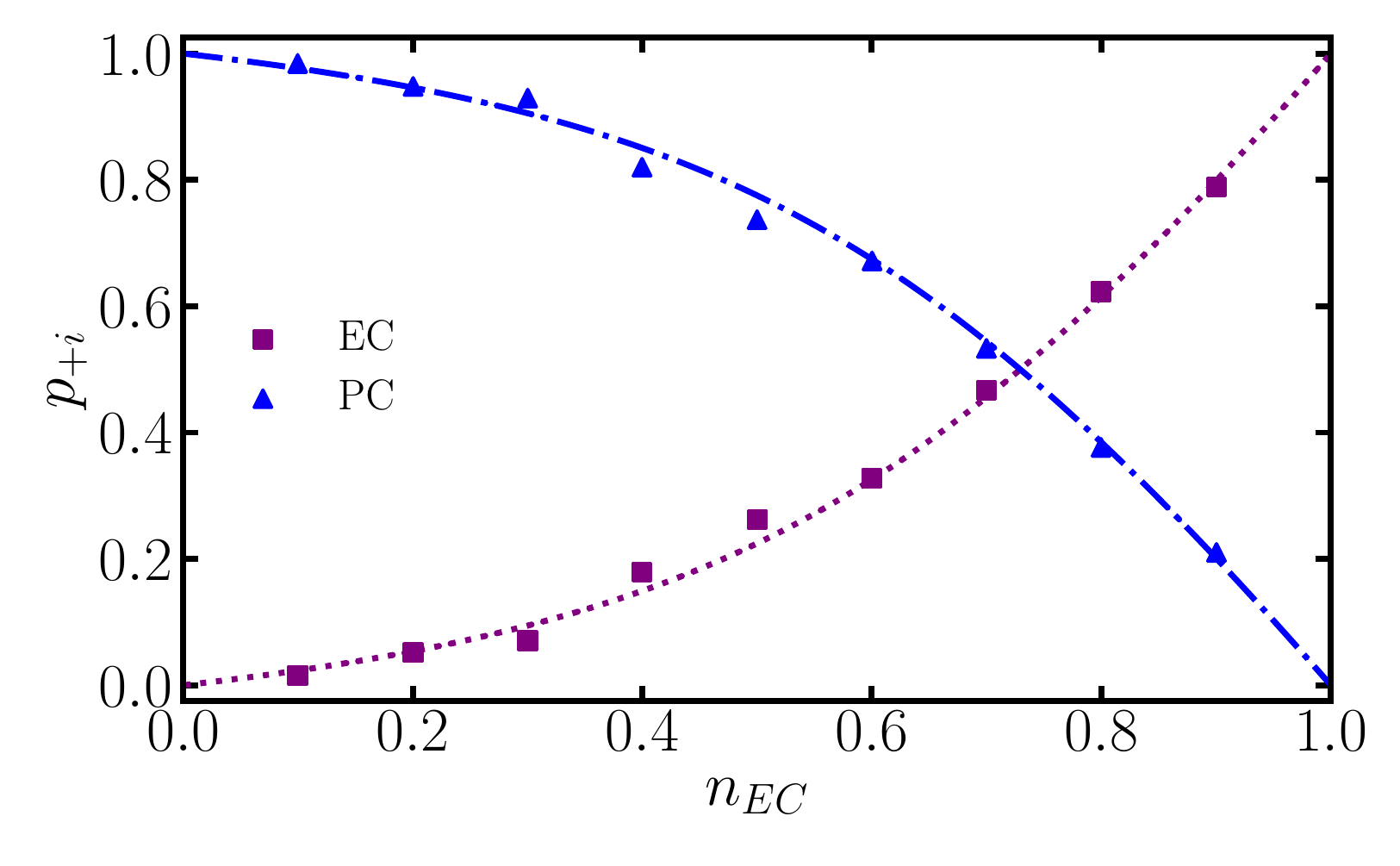}
\includegraphics[width=0.49\textwidth]{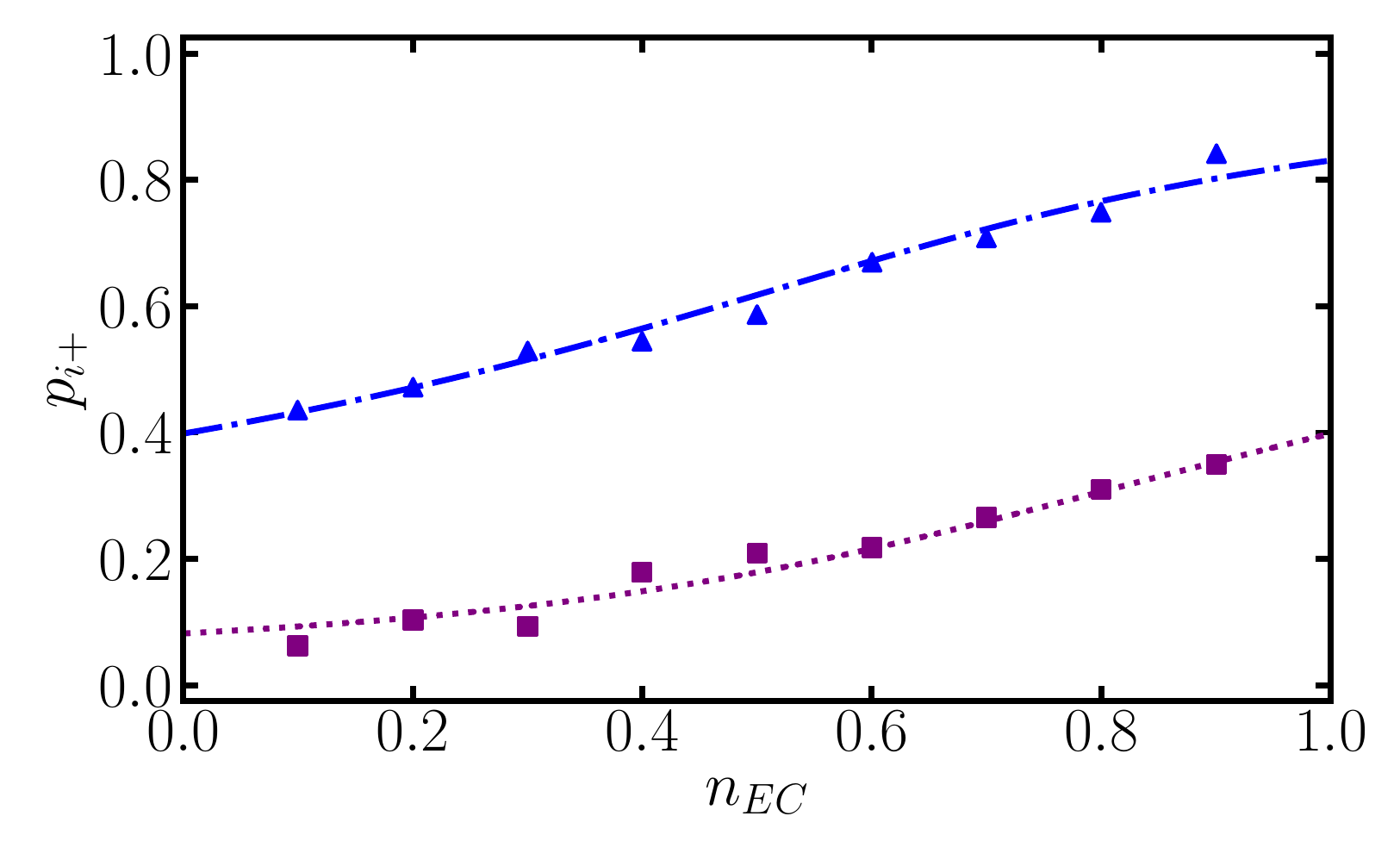}
\caption{Association probabilities as a function of the mole fraction of EC relative to the mole fraction of solvent. Symbols are values directly extracted from simulations. Lines are values calculated from Eqs.~\eqref{eq:prob_1_sticky} and~\eqref{eq:prob_2_sticky} using $f_+ = 4$, $f_{x/y} = 1$ $\tilde{\Lambda} = 0.135$, $\phi_{\pm} = 0.093$ and $\xi_{x/y} = 28.75$.}
\label{fig:simulations_p}
\end{figure} 

To progress further, we make a few simplifications. Firstly, we assume that the volume of EC and PC molecules are the same, and therefore, the volume fraction of LiPF$_6$ is constant (note this is not a necessary assumption, and doesn't change the results significantly, but it is done for convenience). Secondly, the total Li coordination number was consistently 4, which means Li practically had no ``dangling bonds'' because the interactions between Li and the solvents are so energetically favourable. This means the probability of Li binding to solvents is equal to unity, $1 = p_{+x} + p_{+y}$, which we refer to as the ``sticky cation'' approximation~\cite{mceldrew2021ion}. From inspecting Eq.~\eqref{eq:sys2}, with only solvent molecules included in the summation over $i'$, it should be clear there is singular behaviour because the probability of an open cation site, which is $0$ in the sticky cation approximation, sits in the denominator of the equation. This singular behaviour needs to be removed from the system of equations. By dividing the mass action laws between Li and the two different solvents, we arrive at
\begin{equation}
    \dfrac{\Lambda_x}{\Lambda_y} = \Tilde{\Lambda} = \dfrac{p_{x+}(1 - p_{y+})}{p_{y+}(1 - p_{x+})},
    \label{eq:sticky}
\end{equation}

\noindent which is no longer singular. Using this equation, in addition to $1 = p_{+x} + p_{+y}$, permits an analytical solution to the association probabilities 
\begin{align}
    &\psi_+ p_{+x} = \psi_x p_{x+}=\frac{\psi_y-\psi_++\Tilde{\Lambda}(\psi_++\psi_x)}{2(\Tilde{\Lambda}-1)} \nonumber \\
    &-\frac{\sqrt{4\psi_y\psi_+(\Tilde{\Lambda}-1)+[\Tilde{\Lambda}(\psi_x-\psi_+)+\psi_++\psi_y]^2}}{2(\Tilde{\Lambda}-1)},
    \label{eq:prob_1_sticky}
\end{align}
\begin{align}
    &\psi_+ p_{+y} = \psi_y p_{y+} =\frac{\psi_y+\psi_++\Tilde{\Lambda}(\psi_x-\psi_+)}{2(1-\Tilde{\Lambda})} \nonumber \\
    &-\frac{\sqrt{4\psi_y\psi_+(\Tilde{\Lambda}-1)+[\Tilde{\Lambda}(\psi_x-\psi_+)+\psi_++\psi_y]^2}}{2(1-\Tilde{\Lambda})}.
    \label{eq:prob_2_sticky}
\end{align}

\noindent When $\Tilde{\Lambda} > 1$, the associations between Li and EC is favoured over Li and PC ($\Lambda_x > \Lambda_y$), and when $\Tilde{\Lambda} < 1$ the associations between Li and PC is favoured over Li and EC ($\Lambda_y > \Lambda_x$). The other factor in determining the association probabilities is the amount of each species. Therefore, in the sticky cation approximation, we only have a single parameter of the theory which needs to be determined from the MD simulations.

In Fig.~\ref{fig:simulations_p} we show the probabilities computed from the values of $\mathcal{N}_{i}/f_+$ obtained from our MD simulations as a function of $n_{EC}$. Using these probabilities we can compute $\Tilde{\Lambda}$ from Eq.~\eqref{eq:sticky} at each $n_{EC}$ and take the average of the obtained values. We find $\Tilde{\Lambda} = 0.135$, which indicates that PC has more favourable associations with Li than EC (which can also be seen in Fig.~\ref{fig:simulations_p} from $p_{y+} > p_{x+}$ for all $n_{EC}$). Now that $\Tilde{\Lambda}$ has been determined, and there are no other free parameters of the theory, Eqs.~\eqref{eq:prob_1_sticky} and~\eqref{eq:prob_2_sticky} can be used to predict the association probabilities for any value of the variable $n_{EC}$. In Fig.~\ref{fig:simulations_p} we show that these probabilities match the simulated values extremely well. Therefore, our simple, analytical theory can allow us to naturally understand the solvation structure competition of concentrated Li electrolytes. Having understood the association probabilities, we can now turn to the distributions of solvation structures.

\begin{figure}[h]
\centering
\includegraphics[width=0.49\textwidth]{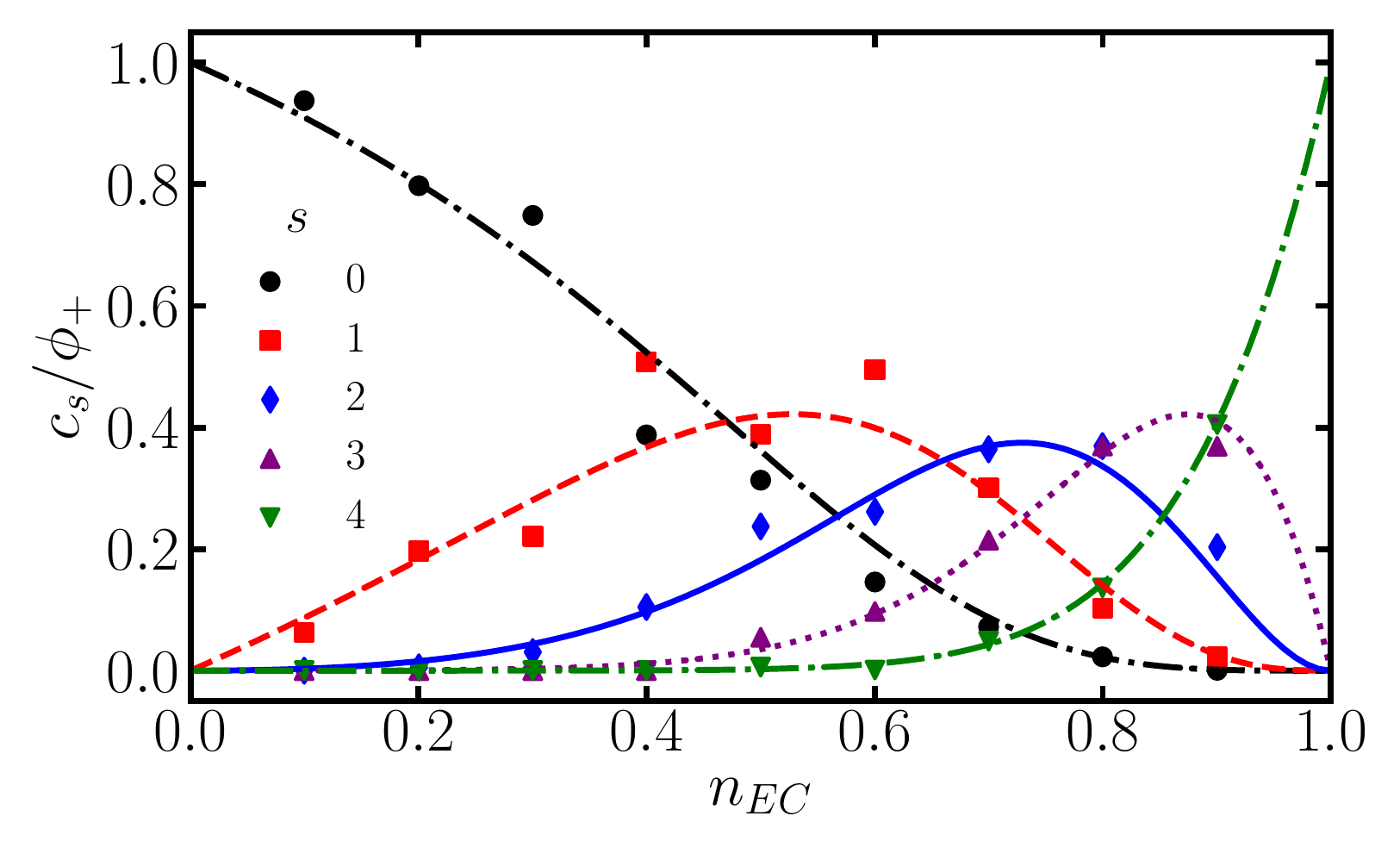}
\caption{Coordination probability of $s$ EC molecules, $c_s/\phi_+$, as a function of $n_{EC}$. The same set of parameters as Fig.~\ref{fig:simulations_p} is used for the lines.}
\label{fig:hist}
\end{figure}

In Fig.~\eqref{fig:hist} we display the probability of $s$ EC solvent molecules in the coordination shell of the Li from the simulations, for $s = 0,...,4$ as a function of $n_{EC}$. For $s = 0$, i.e. Li only being solvated by PC, the probability of this solvation structure is initially close to 1, and it monotonically decreases with increasing $n_{EC}$. Whereas, $s = 1,2,3$, i.e. mixed EC and PC solvation structures, all have a non-monotonic dependence on $n_{EC}$, where they go through a maximum at intermediate $n_{EC}$. Finally, $s = 4$, i.e. only EC associated to Li, monotonically increases with $n_{EC}$, where it is the largest component in the coordination shell at the largest $n_{EC}$. 

To investigate the solvation distribution from the theory, the singular behaviour needs to be removed from Eq.~\eqref{eq:dist} in the sticky cation approximation. Using $1 = p_{+x} + p_{+y}$, $f_+ = s + q$, and Eq.~\eqref{eq:sticky}, we can arrive at
\begin{equation}
    \dfrac{c_{s}}{\phi_+} = \dfrac{f_+!}{s!(f_+ - s)!}p_{+x}^s(1 - p_{+x})^{f_+ - s},
    \label{eq:Cs}
\end{equation}

\noindent which is the probability of $s$ EC solvent molecules in the coordination shell of the Li, with $c_s$ being used as a shorthand for $c_{10sq}$. An equivalent expression for the probability of $q$ PC solvent molecules in the coordination shell of the Li, $c_p$, can be derived, which is also simply $c_p/\phi_+ = 1 - c_s/\phi_+$. Equation~\eqref{eq:Cs} is clearly the number of ways of arranging a solvation shell with $s$ EC and $q$ PC solvent molecules, multiplied by the probability that $s$ solvent molecules are bound to the Li and the probability of the remaining $f_+ - s$ PC molecules being bound to Li, i.e. it is simply a binomial distribution of the two solvent molecules. 

In Fig.~\eqref{fig:hist} we also plot the probability of $s$ EC molecules in the coordination shell of Li from the theory, using the value of $\tilde{\Lambda}$ extracted from the simulations. Overall, we find good agreement between the theory and simulation. The theory allows us to understand this dependence of $c_s/\phi_+$ on $n_{EC}$. For $s=0$, Eq.~\eqref{eq:Cs} reduces to $p_{+y}^{f_+}$, and for $s=f_+=4$, we arrive at $p_{+x}^{f_+}$, which are clearly the probability of $f_+$ PC and EC molecules binding to the Li cation, respectively. As can be understood from Fig.~\ref{fig:simulations_p}, these functions, $p_{+y}^{f_+}$ and $p_{+x}^{f_+}$, are monotonically decreasing and monotonically increasing, respectively. For all values of $s$ between these limits, both $p_{+x}$ and $1 - p_{+x}$ occur in the expressions, which requires that $c_s$ goes to zero at the limits of $n_{EC}$, and that a maximum occurs between these limits. If $\tilde{\Lambda} = 1$, the $s=2$ case would peak at $n_{EC} = 1/2$, but because $\tilde{\Lambda} < 1$ the peak occurs at $n_{EC} > 1/2$ as clusters with PC is favoured and larger concentrations of EC are required to reach this point. This demonstrates that the binomial distribution is skewed based on the relative interactions of the solvents with Li. 

\subsection{Experimental Comparison}

\begin{figure}[h]
\centering
\includegraphics[width=0.49\textwidth]{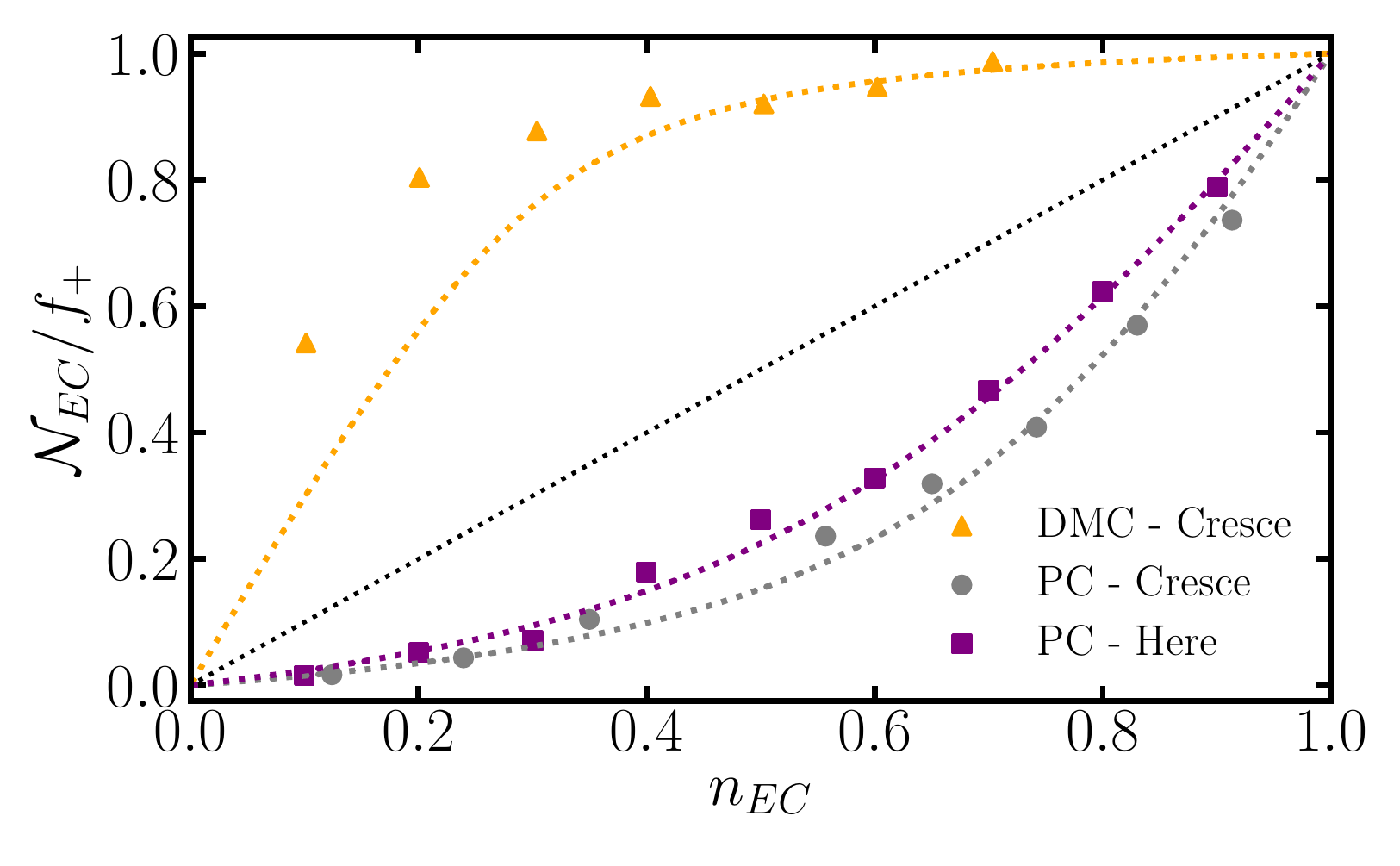}
\caption{Fraction of EC molecules in coordination shell of Li as a function of the mole fraction of EC relative to the mole fraction of solvent, $n_{EC}$. For EC/PC mixtures, the values from Cresce \textit{et al.} in Ref.~\citenum{von2012correlating} and our simulations from the previous section are shown. Also from Ref.~\citenum{von2012correlating}, the EC/DMC mixtures are also displayed. The values of the parameters for the theory were computed from the association probabilities $p_{+x} = \mathcal{N}_{EC}/f_+$. For the experimental values, we take $f_+ = 3$ as it fits the values significantly better, and we find $\tilde{\Lambda} = 0.0978$ for EC/PC and $\tilde{\Lambda} = 26.7$ for EC/DMC. All other values were kept the same as that in Fig.~\ref{fig:simulations_p}}
\label{fig:exp}
\end{figure}

In Ref.~\citenum{von2012correlating} mass spectra of 1M~LiPF$_6$ in mixtures of non-aqueous solvents were measured. As the electrolyte is vaporised and ionised from the liquid state, it is thought that most of the strong associations stay in tact, and therefore, the mass spectrum can provide information of the distribution of solvation structures in the liquid electrolyte. Therefore, the solvation structure of cations is a physical observable which we can directly compare our theory against. In Fig.~\ref{fig:exp} we reproduce Ref.~\citenum{von2012correlating} values for the fraction of EC in the solvation shell, i.e. $\mathcal{N}_{EC}/f_+$, for EC/PC and EC/DMC mixtures, as a function of the mole fraction of EC relative to the mole fraction of total solvent. In addition, we reproduce the values simulated from the previous section. As we know $p_{+x} = \mathcal{N}_{EC}/f_+$ from experiments, assuming the sticky cation approximation, we are able to calculate values of $\Tilde{\Lambda}$ for the experiments, and compare our theory against these values. 

For EC/PC mixtures, the values of $\mathcal{N}_{EC}/f_+$ from simulations and experiments are quite similar. In the experiments, it was noted that only up to 3 solvent molecules coordinating were found. The discrepancy between simulations and experiments as to whether a maximum of 3 or 4 solvent molecules can coordinate lithium in these carbonate blends was discussed in Ref.~\citenum{von2012correlating} and thought to be an attribute of the electrospray ionization technique only retaining the most tightly bound species. Therefore, for the experimental data we use $f_+ = 3$, and we found that $f_+ = 4$ the fitted curves significantly worse than $f_+ = 3$. Based on this functionality, we calculated $\Tilde{\Lambda} = 0.0978$ for the experiments, which is similar to the value obtained from the simulations. This value of $\Tilde{\Lambda}$ indicates that the PC-Li associations are more favourable than the EC-Li associations, which is evident from the curves being below the diagonal dotted line which corresponds to equal numbers of EC/PC in the coordination shell of Li. 

On the other hand, for the EC/DMC mixtures, the EC-Li associations are favoured over the DMC-Li associations, and so $\mathcal{N}_{EC}/f_+$ resides above the diagonal. The value of $\Tilde{\Lambda} = 26.7$ extracted from the experimental data is now larger than 1, demonstrating the more favourable EC-Li associations. Therefore, the value of $\Tilde{\Lambda}$ is a clear, physical parameter which describes the competition between the solvation structure of Li electrolytes. Again, $f_+ = 3$ did a significantly better job at reproducing the data than $f_+ = 4$. 

There are many examples of simulations which show similar trends to those reported in Refs.~\citenum{von2012correlating}. While there can be some differences in the reported experimental or predicted values from simulations, with and without associations between cations-anions, we wish to stress that our theory should be able to capture the behaviour of all these systems, provided the assumptions of the theory are valid. What changes is the exact volume fractions of species, association constants, functionalities, etc.

\section{Beyond Classical Non-aqueous Electrolytes}

In previous sections, we focused on cation solvation in common non-aqueous electrolytes that are studied primarily for use in commercial LIBs. We chose this electrolyte because it has been extensively studied, and the coordination structure is qualitatively agreed upon. While these electrolytes are commonly and successfully used in commercial LIBs, moderately concentrated carbonate blend electrolytes generally do not perform well in many next-generation battery technologies~\cite{li2020new,Tian2021,Kang2022Nav,Piao2022,Cheng2022Sol}. For such applications, there are many emerging classes of electrolyte currently being explored in literature, where our developed theory can be adapted for~\cite{li2020new,Tian2021,Kang2022Nav,Piao2022,Cheng2022Sol,Yu2022Agg,Wang2023soft}. 

In this section, we utilise our theory to understand some ``design principles'' of several classes of next-generation electrolytes currently being investigated. Note we have only picked a few examples to discuss here to demonstrate the theories ability, with an exhaustive comparison being left for future work. We mainly focus on ``toy models'' of the theory for these electrolytes, details of which are shown in the Appendix, and we do not perform simulations to back up these findings, but hope these results inspire further investigation of these systems. For brevity, we will not review in detail why these next-generation electrolytes are of interest or their exact chemistries, but refer readers to where this information can be found. 

\subsection{Localized high-concentration electrolytes}

The design of localized high-concentration electrolytes (LHCE) -- the reader is referred to Ref.~\citenum{cao2021localized} for a review -- starts from finding a promising highly concentrated electrolyte (HCE), typically just a salt dissolved in a solvent that solvates the cation strongly. This HCE is then diluted with a solvent which does not solvate the cation, known as the diluent, in an attempt to preserve the coordination structure of the cation (see Introduction for why this is important). Therefore, the lithium coordination shell -- likely invaded by anions -- resembles one that could typically be obtained at only high salt concentrations in 2-component mixtures. From the perspective of the lithium ions in solution, the electrolyte is ``locally" concentrated.

From our toy model, shown in Appendix~\ref{app:LHCE}, we find that the design principle of LHCE is supported by our theory, under certain approximations. In the sticky cation approximation, the addition of a diluent, which does not interact with any other species, leaves the coordination shell of the cations unchanged, despite the dilution effect. If the sticky cation approximation is not employed, the predictions of the theory sensitively depend on the assumed association constants, and therefore, we do not make any predictions for this case. Using our theory, it should be possible to reveal the necessary components to include in a model that accurately describes these promising electrolytes. 

\subsection{High entropy electrolytes}

Within the chemistries of LHCE's, it has been proposed that by adding more types of solvents, without changing the concentration of salt and the solvation enthalpy, the average cluster size can decreases while retaining roughly the same number of anions in the coordination shell of Li~\cite{Kim2022HEE}. This is thought to occur because of the increased entropy of mixing, similar to high entropy alloys, which is why these systems have been referred to as high entropy electrolytes (HEE)~\cite{Kim2022HEE}. These HEE have been reported to have significantly better transport properties, without diminishing the stability of the electrolyte~\cite{Kim2022HEE}.

The toy model we utilise to investigate this system is shown in Appendix~\ref{app:HEE}. We find that our theory does not support the design principle of HEE. More types of solvent, which interact identically with the cation, leave the average cation-anion cluster size unaltered. In Appendix~\ref{app:HEE}, we have given some possible explanations as to why this could be, but reserve making final, physical conclusions until a detailed comparison can be made between our theory and simulations of these electrolytes. We believe it is essential to perform this comparison to understand the origin of the electrolytes promising transport properties~\cite{Kim2022HEE}.

In addition to Ref.~\citenum{Kim2022HEE}, there are now further examples of electrolytes which claim to be high entropy~\cite{Wang2023HEE1,Wang2023HEE2,Yang2023HEE}. In Appendix~\ref{app:HEE} we further discuss these examples, although we do not specifically develop theories for these cases. We also provide a general discussion of the ideas of high entropy alloys and oxides~\cite{George2019} and what this could mean for electrolytes, and point readers reviews on the role of entropy in self-organised structures~\cite{Sciortino2019ESA,Rocha2020,Frenkel2015}, which will provide insight the role of entropy in concentrated electrolytes, and especially high entropy electrolytes.

\subsection{Chelating Agents}
 
Solvents which can bind to cations from multiple points, i.e. multi-dentate ligands, have been proposed to be beneficial in reversing the negative transference numbers of salt-in-ILs~\cite{molinari2019transport,molinari2019general,mceldrew2020salt}. A limiting case of such electrolytes is when the solvent completely encapsulates the cation, preventing it from binding to any other species, be it anions or other solvent molecules. The ``design principle'' of the addition of encapsulating chelating agents is to break up the cation-anion associations and prevent anions from binding strongly to the cations~\cite{Molinari2020chelating}. Here we shall develop an approach to include these multi-dentate, encapsulating solvents, to see if this principle is supported by our theory.

Our developed theory extension for this type of solvent, shown in Appendix~\ref{aoo:CA}, yields predictions which are consistent with that of Ref.~\citenum{Molinari2020chelating}, i.e. dissociation of larger clusters. Therefore, the design principle of adding chelating agents which encapsulate the cations is supported by our simple theory. Further work should compare how the cluster distribution is altered by the presence of these chelating agents in detail.

\subsection{Soft Solvents}

While mixtures of carbonate solvents have been used in commercial LIB electrolytes for a while, these electrolytes do no perform optimally, and it is known that the solvents are very strongly bound to Li$^+$. Such solvents have high binding energies and dielectric constants, which ensures the solubility of salts such as LiPF$_6$, but causes slow desolvation dynamics, amongst other issues. To overcome this, and related issues, it has been suggested to instead employ ``soft solvents'' which have a fairly low binding energy, but still a modest dielectric constant, in order to reduce the strength of the cation-anion interaction~\cite{Wang2023soft}. With these soft solvents, highly dissociating salts are required, but these mixtures have highly promising transport, interfacial/interphasial properties and temperature operating ranges~\cite{Wang2023soft}.

While we do not need to specifically develop a toy model for soft solvents, as it already falls within the theory presented, we would like to highlight a few points. As we have shown here, the association constants are the physical parameters that describe the association strength, which combines information about the association energy and dielectric constant, as well as the entropy of association~\cite{mceldrew2020corr}. By plotting where different LIB electrolytes are on a graph of $\Lambda_{+-}$ as a function of $\Lambda_{+x}$, the competition of anion and solvent associations for different LIB electrolytes can be characterised. For example, we expect soft solvent systems to have relatively small $\Lambda_{+-}$ and $\Lambda_{+x}$, but commercial LIB electrolytes have a large $\Lambda_{+x}$ and small $\Lambda_{+-}$. The magnitude of the association constants are directly linked to the lifetime of the associations, and therefore, the timescales of association/disassociation~\cite{saika2004effect,Sciortino2008,Tartaglia2009,Hoy2009,corezzi2009connecting,Corezzi2012}.

\section{Conclusions}

In summary, we have developed a theory for the solvation structure and ionic aggregation of non-aqueous electrolytes. We have tested this theory against molecular dynamics simulations of a ``classical'' non-aqueous electrolyte, 1~M LiPF$_6$ in mixtures of carbonates. Overall, we found excellent agreement between our theory and the classical molecular dynamics simulations, and our theory also worked well in comparison to already published experimental results for the competition between EC and PC in the coordination shell of Li$^+$. Confident in the theory, we investigated non-aqueous electrolytes of interest in next generation of LIBs. Overall, it is hoped that this theory will provide a framework for understanding the solvation shell of electrolytes in a more systematic way, and aid the design of next-generation chemistries.

While we have focused on the solvation structure of the cations here, which is known to be crucial in understanding the properties of LIB electrolytes, our theory can be directly used to calculate physiochemical properties of interest, which can quantify the connection between solvation structure and properties of interest. For example, we have developed a consistent treatment of ionic transport based on vehicular motion of species, which could be used to understand conductivity and transference numbers of different mixtures~\cite{mceldrew2020corr,mceldrew2020salt}; we can calculate activity coefficients from our theory~\cite{mceldrew2021ion}, which can provide indications of if the electrolyte is expected to form a stable SEI; and we recently developed the theory to predict the composition of electrolytes at charged interfaces~\cite{Goodwin2022GEL,Goodwin2022IP}, where desolvation of cations could be a result of the breakdown of the cluster distribution at an interface. In our experience, the theory has qualitatively worked for all of these applications, but quantitatively it does best for transference numbers and activity coefficients, with conductivity and double layer properties being more difficult to predict. We believe the value of the theory is its physical transparency that allows a conceptual understanding in these LIB electrolytes.

Looking forward, our theory could be developed to understand the kinetics of SEI formation~\cite{Huang2019SEI,Pinson2013,Das2019} in more detail with explicit account of the solvation shell, and electron transfer reactions of solvated cations at interfaces~\cite{Huang2021ET,Fraggedakis2021}, where the reorganisation energy could be linked to the free energy of forming those clusters. Here we focused on non-aqueous electrolytes, and previously we have studied ionic liquids~\cite{mceldrew2020corr}, water-in-salt electrolytes~\cite{mceldrew2021ion} and salt-in-ionic liquids~\cite{mceldrew2020salt}, but the theory is quite general, and could be applied to a wide variety of electrolytes, from high concentration to low concentration (provided long-ranged electrostatics are introduced~\cite{mceldrew2020theory}), for various temperatures~\cite{mceldrew2020corr}, and for various types of solvent (aqueous and non-aqueous) and salt compositions, provided Cayley tree clusters are formed.

The theory we present here is simple, but we believe it acts as a good starting point for further work. One of the main limiting assumptions of the theory is the formation of Cayley trees, with only 2 species that can form the backbone of a percolating network. In reality, this assumption only holds for specific electrolytes, typically only with chaotropic species. Fortunately, LIB electrolytes mainly appear to fall within this category, but certain high entropy electrolytes, or electrolytes with only kosmotropic species break our assumptions. Therefore, we believe it is important to further develop theories~\cite{Stillinger1963,Teixeira2019,Andersen1974} and perform simulations beyond the assumptions here~\cite{Doye1999,choi2014ion,choi2015ion,choi2015ion3,choi2017ion}.

\section{Acknowledgements}

We thank Nicola Molinari, Julia Yang, Sang Cheol Kim and Jingyang Wang for stimulating discussions. We acknowledge the Imperial College London Research Computing Service (DOI:10.14469/hpc/2232) for the computational resources used in carrying out this work. MM and MZB acknowledge support from a Amar G. Bose Research Grant. The work at Harvard was supported by the Department of Navy award N00014-20-1-2418 issued by the Office of Naval Research.

\appendix
\renewcommand{\thefigure}{A\arabic{figure}}
\setcounter{figure}{0} 
\renewcommand{\theequation}{A\arabic{equation}}
\setcounter{equation}{0} 

\section{Localized high-concentration electrolyte}
\label{app:LHCE}

Our toy model for localized high-concentrated electrolyte (LHCE) first fixes a relative composition of highly concentrated electrolyte (HCE) with cations, and anions and solvent molecules which bind to the cations. All of these components are then equally diluted. This additional co-solvent which dilutes the HCE, making the LHCE, is introduced as any other solvent in the theory, but with an association constant of $0$. Therefore, no additional mass action or conservation of association equations need to be introduced, and all that changes is the new volume fractions of the cation, anion and solvent which are given by $\phi_i^d = \phi_i(1 - \phi_d)$, where $\phi_d$ is the volume fraction of the diluent and $\phi_i^d$ is the new volume fraction of each diluted species. 

Let us first look at the predictions from the ``sticky cation'' approximation. For the anion and a single solvent in the coordination shell of the cation, we have
\begin{equation}
    \dfrac{\Lambda_-}{\Lambda_x} = \Tilde{\Lambda} = \dfrac{p_{-+}(1 - p_{x+})}{p_{x+}(1 - p_{-+})},
    \label{eq:sticky_SM}
\end{equation}

\noindent and $1 = p_{+-} + p_{+x}$, permits an analytical solution for the association probabilities 
\begin{align}
    &\psi_+ p_{+-} = \psi_- p_{-+}=\frac{\psi_x-\psi_++\Tilde{\Lambda}(\psi_++\psi_-)}{2(\Tilde{\Lambda}-1)} \nonumber \\
    &-\frac{\sqrt{4\psi_x\psi_+(\Tilde{\Lambda}-1)+[\Tilde{\Lambda}(\psi_--\psi_+)+\psi_++\psi_x]^2}}{2(\Tilde{\Lambda}-1)},
    \label{eq:prob_1_sticky_SM}
\end{align}
\begin{align}
    &\psi_+ p_{+x} = \psi_x p_{x+} =\frac{\psi_x+\psi_++\Tilde{\Lambda}(\psi_--\psi_+)}{2(1-\Tilde{\Lambda})} \nonumber \\
    &-\frac{\sqrt{4\psi_x\psi_+(\Tilde{\Lambda}-1)+[\Tilde{\Lambda}(\psi_--\psi_+)+\psi_++\psi_x]^2}}{2(1-\Tilde{\Lambda})}.
    \label{eq:prob_2_sticky_SM}.
\end{align}

In the above equations, the replacement of $\psi_i$ for $\psi_i^d$ can be made. It is clear that the factor $(1 - \phi_d)$ which multiplies all original volume fractions of species completely cancels from the expressions. Therefore, for a fixed $\Tilde{\Lambda}$, within the sticky cation approximation, there is no effect from diluent on the association probabilities, and the coordination shell of the cation remains identical to the HCE it is derived from.

To further understand LHCE, let us investigate a mass action law where $1 > \sum_{i'}p_{+i'}$, i.e. not the sticky cation approximation. If the association probabilities are to remain unchanged by the addition of the diluent, then $\Lambda_i\Gamma_i$ must remain unchanged. The value of $\Gamma_i$ should, however, decrease with increasing volume fraction of diluent. Therefore, there must be a corresponding increase in $\Lambda_i$ to compensate, to keep $\Lambda_i\Gamma_i$ constant. When taking the ratio of the association constants in the sticky cation approximation this increase in the $\Lambda_i$'s was hidden.

It is known that the coordination structure of the cation does change~\cite{cao2021localized}, which is perhaps not surprising, as in reality $\Tilde{\Lambda}$ (or the $\Lambda_i$'s) will not correspondingly increase. It is thought that the diluent can interact with the solvent molecules as well as change the dielectric constant of the mixture resulting in an increase of anion molecules in the cation coordination shell~\cite{cao2021localized}. 

One can account for short-ranged interactions between the solvent and diluent through a regular solution interaction, similar to how we treated the IL cations in salt-in-ILs~\cite{mceldrew2020salt}. This will tend to remove the solvent molecules from the cations solvation shell, which should exponentially depend on the volume fraction of the diluent. By computing the concentration dependence of the association constant on the diluent volume fraction, it should be clear exactly how the interaction appears.

On the other hand, the diluent changes the dielectric constant of the mixture and the concentration of associating species which, presumably, alters the association constants in a different way. Predicting the exact changes which will happen is difficult, but by performing simulations of specific LHCE and extracting the association probabilities and constants, one can further understand the simulations from the developed theory.

\section{High entropy electrolytes}
\label{app:HEE}

Our toy model for the high entropy electrolytes (HEE) of Ref.~\citenum{Kim2022HEE} assumes a fixed volume fraction of salt, with the cation-anion association constant fixed. The total volume fraction of solvent also remains constant, but its composition can be divided up between multiple different solvents. For now, let us assume these solvents can solvate the cation, and there is no diluent present. These solvents are essentially only different in the sense that they are labelled differently, as we take their volumes ($\xi_1 = \xi_{2,2} = \xi_{2,1}$) and association constants with the cations to be identical ($\Lambda_1 = \Lambda_{2,2} = \Lambda_{2,1}$), where the index $1$ is used for the single solvent, and $2,1$ and $2,2$ is used to denote the solvents in the system with two solvents. Therefore, the only difference between a system with 1 solvent (with $\phi_1$) and another with 2 solvents (with $\phi_1 = \phi_{2,1} + \phi_{2,2}$) is the entropy of the systems. We include the entropy of mixing, configurational entropy and conformational entropy of clusters, as shown in the Theory Section of the main text.

Furthermore, let us first focus on a limiting case where the solvent cation interactions are extremely strong, $\Lambda_i \gg 1$, with the cation-anion association constant much smaller $\Lambda_i \gg \Lambda_-$, and when there are more cation association sites than solvent association sites, i.e. $c_+f_+ > \sum_ic_i$. In such a case, the probability of solvent associating to cation is, of course, $p_{i+} = 1$. From the conservation of associations, the cation association probabilities to the solvents are $p_{+i} = \psi_i/\psi_+$. For the single solvent system we have $p_{+1} = \psi_1/\psi_+$, and for the two solvent system we have $p_{+2,2} = \psi_{2,2}/\psi_+$ and $p_{+2,1} = \psi_{2,1}/\psi_+$. Clearly, then, we find that the total probability of a cation being associated to a solvent molecule remains the same between 1 and 2 types of solvent molecules, with the relationship $p_{+1} = p_{+2,1} + p_{+2,2}$ holding. In fact, as we found from numerically solving the mass action laws, this relationship holds irrespective of the employed association constants (provided the solvent-cation association constants are the same) and volume fractions. 

With the previously derived results for $p_{+i}$, it is clear that the cation-anion mass action law, as generally given by
\begin{equation}
    \Lambda_-\Gamma_- = \dfrac{p_{+-}p_{-+}}{\left(1 - p_{+-} - \sum_{i}p_{+i}\right)(1 - p_{-+})},
\end{equation}

\noindent remains unchanged by the presence of more types of solvents which interact with the cation identically. Unsurprisingly, then, the cation-anion probabilities remain the same, independent of the number of ``different'' solvents in the system. As the cations and anions form the back-bone of the ``alternating ionic co-polymer'' of Cayley-tree-like clusters, if their probabilities of association do not change, then the cation-anion cluster distribution, after summing out the solvent indexes, does not change. Moreover, the percolation threshold is determined by $(f_+ - 1)(f_- - 1) = p_{+-}p_{-+}$ for Cayley tree clusters, which also cannot change from adding more types of ``different'' solvents.

We will explicitly show how the cluster distribution is not effected by the number of different solvents, again by considering the case of 1 solvent and 2 solvents. For clarity of notation, we take $\Lambda = \Lambda_1 = \Lambda_{2,2} = \Lambda_{2,1}$. The cluster distribution for 1 solvent is generally given by 
\begin{equation}
    c_{lms} = \frac{W_{lms}}{\Lambda_{-}}\left(\psi_{l}\Lambda_{-} \right)^{l} \left(\psi_{m}\Lambda_{-} \right)^{m}(\psi_{s}\Lambda)^s.
    \label{eq:dist_1_SM}
\end{equation}

\noindent Let us assume that there is some combination of $l$ and $m$ such that there are only $n$ cation association sites remaining. Therefore, we arrive at
\begin{equation}
    \dfrac{c_{ns}}{c_{n}} = \dfrac{n!}{s!(n-s)!}(\psi_{s}\Lambda)^s,
    \label{eq:clms_SM}
\end{equation}

\noindent where $c_{n}$ is used as a short-hand for the cluster distribution of this cation-anion combination with $n$ cation association sites remaining.

Similarly, the cluster distribution for two solvents is
\begin{equation}
    c_{lmpq}=\frac{W_{lmpq}}{\Lambda_{-}}\left(\psi_{l}\Lambda_{-} \right)^{l} \left(\psi_{m}\Lambda_{-} \right)^{m}(\psi_{p}\Lambda)^p(\psi_{q}\Lambda)^q,
    \label{eq:dist_2_SM}
\end{equation}

\noindent and again considering $n$ cation association sites being free, we have
\begin{equation}
    \dfrac{c_{npq}}{c_{n}} = \dfrac{n!}{p!q!(n-p - q)!}(\psi_{p}\Lambda)^p(\psi_{q}\Lambda)^q.
    \label{eq:clmsq_SM}
\end{equation}

\noindent Given the probabilities $p_{i+}$ are not altered, and that the volume fraction of the solvent remains constant, it is clear that the following relationship holds for any $s$
\begin{equation}
    c_{ns} = \sum_{pq}^{s = p + q}c_{npq}.
\end{equation}

\noindent This is simply a binomial distribution of the two solvents in place of the single solvent, which does not alter how much solvent is associated to the clusters. While we have shown this for only two solvent here, it is expected that it generalises to any number of solvents, with there being a multinomial instead of a binomial distribution.

Another example of a HEE could be from keeping the solvating solvents the same, but increasing the number of diluent solvents. From the previous section on LHCE, as these diluents do not associate to the cation ($\Lambda = 0$), if they interact with the solvnets identically (say through the same regular solution term $\chi$), and have the same dielectric constant, it is clear that more diluents should not alter the cation-anion association probabilities or the cation-solvent probabilities.  

We can conclude, then, that the design principle of HEE is not supported by our theory. We find that, for solvents which interact with the cation (and other solvents) identically, more types of solvents does not alter the cation-anion cluster distribution or percolation point, provided the cation-anion association constant is fixed.

These electrolytes are, however, advantageous for some reason, as their transport properties clearly show~\cite{Kim2022HEE}. There could be two possible reasons why our theory does not support the assumed principle of HEE. One possibility is that the fundamental assumptions of our theory (Cayley tree clusters, neglecting long-ranged electrostatics, the free energy of formation of a cluster, mono-dentate solvent molecules, etc.) are flawed in some way and cannot be applied to HEE. The other possibility is that, while the solvents interact with the cation in a similar way, they are never \textit{exactly} the same, and the changes to the properties of the electrolyte occur because of the slight differences in their association constants (or functionalities, volumes, etc.), instead of the proposed increase in the entropy of mixing. Only by a detailed comparison between the proposed toy model and atomistic simulations can the origin of the discrepancy be unveiled. In Appendix~\ref{app:report}, a method for systematically performing such comparisons of solvation and aggregation behaviour is outlined.

In addition to Ref.~\citenum{Kim2022HEE}, there are now further examples of ``high entropy electrolytes''. In Ref.~\citenum{Yang2023HEE}, Yang \textit{et al.} combined LiCl and ZnCl$_2$ in water at various compositions, and found a 2:1 ratio with 9 water molecules per formula unit yielded an optimal transport and stability compromise, performing significantly better than water-in-salt~\cite{Suo2015}. This occurs because the Li$^+$ coordination shell is mostly water, with Zn$^{2+}$ being consistently coordinated by 4Cl$^-$ anions. The authors argued this electrolyte was distinct from ``water-in-bi-salt'', because of the observed properties being distinct from previous water-in-bi-salt systems~\cite{Reber2021,Yamada2016}. We, however, disagree with this conclusion and believe the difference in the behaviour occurs from differences of interactions of the cations with water and chloride~\cite{Yang2023HEE}. In previous water-in-bi-salt systems~\cite{Reber2021}, there was typically only one strongly coordinating cation, with the other being a weakly interacting ionic liquid cation. This meant the water and anion are in competition for the coordination sites of the cation. In Ref.~\citenum{Yang2023HEE}, however, two strongly coordinating cations are used, and their preferential solvation behaviour and 2:1 ratio assists Li$^+$ being solvated by water and Zn$^{2+}$ by Cl$^-$. We believe this system can be modelled by a combined and modified version of our theories for salt-in-ionic liquids~\cite{mceldrew2020salt} and water-in-salt electrolytes~\cite{mceldrew2021ion}.


In Ref.~\citenum{Wang2023HEE2}, the authors investigated the liquid-solid phase behaviour of mixtures of Li$^+$ salts with numerous anions. Overall, it is found when 5 anions are present, there is a suppression of precipitation. It is claimed this occurs because the larger number of different anions in the coordination shell of Li$^+$. While we cannot explicitly develop a theory for this example, as there are many species with $f_i >1$, this conclusion would very naturally follow from our theory if this limitation was overcome, provided Li$^+$ interacts with all anions with a similar free energy of association. Moreover, linking the suppression of precipitation from disordered coordination environments is conceptually similar to engineering the phase behaviour in high entropy alloys/oxides. We believe in this electrolyte kosmotropic, ordered clusters of 2 species should be suppressed is favour of chaotropic, branched aggregates with many different species. In addition, it was also found transport properties were improved, but no link was suggested to the exact coordination environment, other than it being disordered. Overall, we believe extensions of our theory will support the conclusions made for this electrolyte being high entropy, but further analysis of the simulation results will be required to completely understand this promising system.

In high entropy alloys and high entropy oxides, the crucial requirements are that a large number of similar components exist, often at least 5, and that the difference in how these species interact with other species is similar, which motivates neglecting interaction terms and only retaining the entropy of mixing~\cite{George2019}. We believe these two requirements, at least $\sim$5 components of cations/anions/solvents/diluents and these $\sim$5 components interacting similarly with the other components, are the minimum requirements for an electrolyte to be high entropy. In addition, in LIB electrolytes, presumably only one active cation is desired, which means only multiple anions/solvents/diluents are potentially more promising avenues. In Ref.~\citenum{Kim2022HEE} they showed that the solvents had a similar solvation enthalpy, but only 3 different coordinating solvents were considered, which might not conceptually be thought of as high entropy (at least according to high entropy alloys/oxides~\cite{George2019}, but in electrolytes the number of components that make it high entropy could be smaller than that of alloys/oxides). In Ref.~\citenum{Yang2023HEE} again only two cations were considered, and they interacted with water and the chloride ions very differently, which means both conditions are not satisfied. In Ref.~\citenum{Wang2023HEE2}, 5 different anions were investigated, but it was not explicitly shown that they have a similar interaction strength with the other components, although their reported clusters suggest they could be similar. Thus Ref.~\citenum{Wang2023HEE2} should satisfy both of the requirements for a high entropy electrolyte.

It is also crucial to note that the role of entropy in self-assembling systems is very different to, and significantly more complicated than, high entropy alloys/oxides. In high entropy alloys/oxides there is essentially only the entropy of mixing. In the theory presented here, which builds on previous works on self-assembled systems, there is the entropy of mixing all clusters, but also the combinatorial and configurational entropy of the clusters. For free species the combinatorial entropy is 0, but when there are large species with many ways of arranging them, the entropy substantially increases. Therefore, entropy actually drives the formation of large clusters. We wish to point the reader to Refs.~\citenum{Sciortino2019ESA,Rocha2020,Frenkel2015} for further discussion of these concepts in self-assembled systems, as these ideas will be useful to understand high entropy electrolytes.

\section{Chelating Agents}
\label{aoo:CA}

In this section, we outline modifications to the theory that allow chelating agents to be included. This is the starting point for the inclusion of solvent molecules which are multi-dentate, but that do not completely encapsulate a cation. We do not solve this problem in its full complexity, but simply outline the principles, leaving an in-depth analysis for future work.

The chemical equilibrium between free species and the higher ranking clusters determines the cluster distribution, and the free cation concentration is one of the key variables in this balance. If there are anions/solvents and another solvent which can encapsulate the cation, the free cation concentration is given by the probability of not being bound to the anions/solvents and the encapsulating solvent, as seen through
\begin{equation}
    \phi_{100} = \phi_+p_{f}p_{c},
\end{equation}

\noindent where the probability of not being bound to any anions or singly solvating molecules given by
\begin{equation}
    p_{f} = (1 - \sum_ip_{+i})^{f_+},
\end{equation}

\noindent and the probability of not being encapsulated by one of the chelating agents is given by
\begin{equation}
    p_{c} = 1 - p_{+c}.
\end{equation}

\noindent Crucially, the cation behaves as if it has a functionality of 1 when binding to the encapsulating solvent molecule. 

The free cation concentration can be recast into a more familiar form
\begin{equation}
    \phi_{100} = \phi_+^cp_{c},
\end{equation}

\noindent where there is a renormalised volume fraction of cations which are not bound to anions/solvent
\begin{equation}
    \phi_+^c = \phi_+(1 - \sum_ip_{+i})^{f_+}.
\end{equation}

\noindent Equivalently, the free cation volume fraction can be written as
\begin{equation}
    \phi_{100} = \phi_+^ip_{f},
\end{equation}

\noindent with a renormalised cation volume fraction without any associations to the encapsulating solvent
\begin{equation}
    \phi_+^i = \phi_+(1 - p_{+c}).
\end{equation}

\noindent The important point to note is that the renormalised volume fractions of cations can be treated as unknowns which need to be found.

From these observations, a set of modified mass action laws and conservation of associations can be stated. For the encapsulating solvent, we have for the mass action law
\begin{equation}
    \Lambda_c \Gamma_c = \dfrac{p_{+c}p_{c+}}{(1 - p_{+c})(1 - p_{c+})},
     \label{eq:mass_c}
\end{equation}

\noindent and for the conservation of associations 
\begin{equation}
    \Gamma_c = \psi_+^cp_{+c} = \psi_cp_{c+},
     \label{eq:con_c}
\end{equation}

\noindent where $\psi_+^c = \phi_+^c/\xi_+ = \phi_+(1 - \sum_ip_{+i})^{f_+}/\xi_+$, with $i \neq c$. 

The mass action laws for the cation associations to the anions/solvents are similar to the case before
\begin{equation}
    \Lambda_i\Gamma_i^c = \dfrac{p_{+i}p_{i+}}{(1 - \sum_ip_{+i})(1 - p_{i+})},
    \label{eq:mass_i}
\end{equation}

\noindent and the conservation of associations
\begin{equation}
    \Gamma_i^c = \psi_+^ip_{+i} = \psi_ip_{i+},
     \label{eq:con_i}
\end{equation}

\noindent where $\psi_+^i = \phi_+^ip_{i+}/\xi_+ = \phi_+(1 - p_{+c})/\xi_+$. 

Now, Eqs.~\eqref{eq:mass_c}-\eqref{eq:con_i} need to be solved to obtain the association probabilities, and therefore, the cluster distribution. Either this can be done numerically all together. Alternatively, one can solve Eqs.~\eqref{eq:mass_c} and \eqref{eq:con_c} together to obtain the association probabilities $p_{+c/c+}$ as a function of $\phi_+^c$, and Eqs.~\eqref{eq:mass_i} and \eqref{eq:con_i} to obtain association probabilities $p_{+i/i+}$ as a function of $\phi_+^i$. These resulting probability equations, which are functions of $\phi_+^c$ and $\phi_+^i$, need to be self-consistently solved to find the consistent values of the association probabilities. 

Instead of solving the set of equations in its full complexity, we only study a single limiting case. We take $\Lambda_c \gg \Lambda_i$. Therefore, practically all of the chelating agents are bound to the cations. This means $p_{c+} = 1$ and $p_{+c} = \phi_c/\phi_+ < 1$ from the conservation of associations. Therefore, $\phi_+^i = \phi_+ - \phi_c$, which is a known quantity, and can be used in Eqs.~\eqref{eq:mass_i} and \eqref{eq:con_i} to obtain association probabilities $p_{+i/i+}$.

Let us again study the limit of the sticky cation approximation, given by Eqs.~\eqref{eq:prob_1_sticky_SM} and~\eqref{eq:prob_2_sticky_SM}, but where one of the solvents is replaced by anions. Again, let us start from a HCE ($1 = \phi_+ + \phi_- + \phi_x$) and add the chelating agent, and see how the association probabilities change between these two systems. The chelating agent reduces the overall volume fraction of all other species, but from LHCE we know that this does not alter the association probabilities, provided $\Tilde{\Lambda}$ is constant. Relative to the HCE without the chelating agent, the new volume fraction of anions is given by $\phi_-' = \phi_-(1 - \phi_c)$ and the new volume fraction of solvent by $\phi_x' = \phi_x(1 - \phi_c)$, while the new volume fraction of cations is give by $\phi_+' = \phi_+(1 - \phi_c) - \phi_c$, assuming all the chelating agent binds to the cation. 

We find that (assuming equal volumes and numbers for everything $\phi_+ = \phi_- = \phi_x$ and $c_+ = c_- = c_x$, $f_+ = 4$, $f_- = 3$, which is typical) for solvent-cation associations favoured over the anion-cation associations, $\Tilde{\Lambda} = \Lambda_-/\Lambda_x < 1$, the addition of chelating agent causes a reduction in both $p_{+-}$ and $p_{-+}$, which clearly reduces $p_{+-}p_{-+}$ and the extent of large clusters. There is an increase in $p_{+x}$ to compensate for the decreased $p_{+-}$ (because of the sticky cation approximation), and a decrease in $p_{x+}$ from the conservation of associations. Therefore, the chelating agent breaks up the aggregates. On the other hand, for $\Tilde{\Lambda} = \Lambda_-/\Lambda_x > 1$, the addition of the chelating agents increases $p_{+-}$ but decreases $p_{-+}$, and results in both $p_{+x}$ and $p_{x+}$ decreasing. The increase in $p_{+-}$ is small in comparison to the decrease in $p_{-+}$, which means $p_{+-}p_{-+}$ decreases and breaks up larger clusters in favour of smaller ones.

Therefore, it appears that it is always appears to be advantageous to introduce a chelating agent to break up large clusters~\cite{molinari2019transport,molinari2019general}, and the design principle of adding these molecules is supported by our theory~\cite{Molinari2020chelating}.

\section{Consistent Reporting of Associations}
\label{app:report}

Molecular dynamics simulations are often the method of choice for understand non-aqueous electrolytes, but the quantities which are reported from such simulations often differ, making it difficult to compare two studies. For example, sometimes the percentages of $0, 1, 2, 3, ...$ anions coordinated to the cation is shown, or the percentage of free/bound solvent, pictures of percolating aggregates, or the numbers/frequency of certain coordination environments are reported. This inconsistency in reporting the associations makes it difficult to systematically compare studies, and often not enough information is provided to fully characterise the associations. Here we outline a convention for reporting associations, as the developed theory provides a framework to make reporting such results more systematic.

As is usually done in some form already, the ensemble average coordination number of species $j$ associated to species $i$, $\mathcal{N}_{ij}$, should be stated (note the index $i$ was dropped in the main text, as only the coordination shell of cations was considered). This information can be used to calculate/infer the functionality of each species $f_i$, as the sum of the average coordination numbers should be a maximum of $f_i$ (within some tolerance). The $\mathcal{N}_{ij}$ can also be used to calculate the association probabilities using the $f_i$'s through $p_{ij} = \mathcal{N}_{ij}/f_i$, or from the conservation of associations and the volume fractions. While this is a natural starting point, only reporting the directly associated species does not always uniquely characterise all the clusters, and it is necessary to calculate further quantities.

A requirement of our theory is that only 2 species can have functionalities larger than 1, with all other species only being allowed functionalities of 1. This requirement is a reflection of the fact that if there are 3 species with functionalities larger than 1, the Cayley tree clusters are no longer uniquely defined by their rank. So far we have only worked with electrolytes within this requirement, but it is clear that the promising properties of high entropy electrolytes will mean studying compositions outside of this requirement.

For the case of a maximum of 2 species with a functionality larger than 1, if the electrolyte (almost) exclusively forms Cayley tree clusters, the clusters are uniquely defined by their rank and the cluster distribution $c_{lmsq...}$ can be reported. To display the cluster distribution of this case, 2D maps of the $c_{lm} = \sum_{sq...}c_{lmsq...}$ should be displayed, where $l$ and $m$ are the species which can percolate ($f_i > 1$), and $sq...$ have $f_i = 1$. In addition, as single cation species are presumably of most interest (as aggregates with more than 1 cation are typically large and immobile), 2D concentration plots over different combinations of anion/solvents or solvent/solvent would clearly show these clusters of most interest. If the inequality $p_{ij}p_{ji} < (f_j - 1)(f_i - 1)$ is held, the system is in the pre-gel regime. If $p_{ij}p_{ji} \geqslant (f_j - 1)(f_i - 1)$, the system is in the post-gel regime, and in addition to the cluster distribution being specified, the fraction of each species in the gel must also be stated.

Crucially, to test if Cayley tree clusters are formed, the cluster bond density can be calculated through the number of associations in a cluster divided by the number of species in that cluster. For Cayley trees, the cluster bond density is $1-1/N$, where $N$ is the number of species in the cluster. For very large clusters, the cluster bond density of Cayley trees saturates at 1. If the cluster bond density (substantially) exceeds $1-1/N$, then there are (significant) intra-cluster loops, indicating more ordered aggregates than Cayley trees are formed. If this is the case, the cluster distribution $c_{lmsq...}$ no longer uniquely defines the system, and for each cluster of rank $lmsq...$ there exists another distribution on how this cluster is connected (with different numbers of associations to different species), which must be specified to uniquely define the system. While it is unclear exactly how these results should be reported from our theory, we believe that reporting results just based on the rank of the cluster still provides valuable information, following the previously suggested plots, and would be a good starting point to quantifying the clusters.

If there are more than 2 species with functionalities larger than 2, even if Cayley tree clusters are formed, the rank of the cluster does not uniquely define the cluster. Therefore, how the clusters are connected must always be reported for this case. It is still important to calculate the cluster bond density, as this information provides insight into how ordered clusters are. Again, as it is unclear from the theory exactly how to present the results of these clusters, we suggest that the convention of 2 species with $f_i > 1$ is followed, but where all of the relevant 2-species combinations is displayed on a 2D plot. Similarly, the volume fractions of each species in the gel should be reported, if present, although the percolation condition for this case is not known.

\bibliography{main.bib}

\end{document}